\documentclass[twocolumn]{aastex63}
\newcommand{\psj}{Planet. Sci. J.}

\submitjournal{PSJ}

\graphicspath{{./}{figures/}}

\usepackage{tikz}
\usetikzlibrary{arrows,shapes,positioning,shadows,trees}

\tikzset{
  basic/.style  = {draw, text width=4cm, drop shadow, font=\sffamily, rectangle},
  root/.style   = {basic, rounded corners=2pt, thin, align=center, fill=blue!10},
  level 2/.style = {basic, rounded corners=6pt, thin,align=center, fill=pink!40, text width=11em},
  level 3/.style = {basic, thin, align=center, fill=white!60, text width=8em}
}

\usepackage[utf8]{inputenc}
\usepackage{dirtytalk}
\usepackage{xcolor}

\begin{document}

\title{A pre-flyby view on the origin of asteroid Donaldjohanson, a target of the NASA Lucy mission}

\correspondingauthor{Simone Marchi}
\email{marchi@boulder.swri.edu}

\author[0000-0003-2548-3291]{Simone Marchi}
\affiliation{Department of Space Studies, Southwest Research Institute, 1050 Walnut St., Suite 300,
             Boulder, CO 80302, United States}
\author[0000-0002-6034-5452]{David Vokrouhlick\'y}
\affiliation{Astronomical Institute, Charles University, V Hole\v{s}ovi\v{c}k\'ach 2,
             CZ 18000, Prague 8, Czech Republic}
\author[0000-0002-4547-4301]{David Nesvorn{\' y}}
\affiliation{Department of Space Studies, Southwest Research Institute, 1050 Walnut St., Suite 300,
             Boulder, CO 80302, United States}
\author[0000-0002-1804-7814]{William F. Bottke}
\affiliation{Department of Space Studies, Southwest Research Institute, 1050 Walnut St., Suite 300,
             Boulder, CO 80302, United States}
\author[0000-0003-4914-3646]{Josef \v{D}urech}
\affiliation{Astronomical Institute, Charles University, V Hole\v{s}ovi\v{c}k\'ach 2,
             CZ 18000, Prague 8, Czech Republic}
\author[0000-0001-5847-8099]{Harold F. Levison}
\affiliation{Department of Space Studies, Southwest Research Institute, 1050 Walnut St., Suite 300,
             Boulder, CO 80302, United States}

\begin{abstract}
  The NASA Lucy mission is scheduled to fly-by the main belt asteroid (52246) Donaldjohanson on April 20, 2025. Donaldjohanson (DJ hereafter) is a member of the primitive (C-type class) Erigone collisional asteroid family located in the inner main belt in proximity of the source regions of asteroid (101955)~Bennu and (162173)~Ryugu, visited respectively by OSIRIS-REx and Hayabusa2 missions. In this paper we provide an updated model for the Erigone family age, and discuss DJ evolution resulting from non-gravitational forces (namely Yarkovsky and YORP), as well as its collisional evolution. We conclude the best-fit family age to be $\sim 155$~Myr, and that, on such timescales, both Yarkovsky and YORP effects may have affected the orbit and spin properties of DJ. Furthermore, we discuss how the NASA Lucy mission could provide independent insights on such processes, namely by constraining DJ shape, surface geology and cratering history.
\end{abstract}


\section{Introduction} \label{intro}
\begin{figure}[t!]
 \begin{center}
 \includegraphics[width=0.47\textwidth]{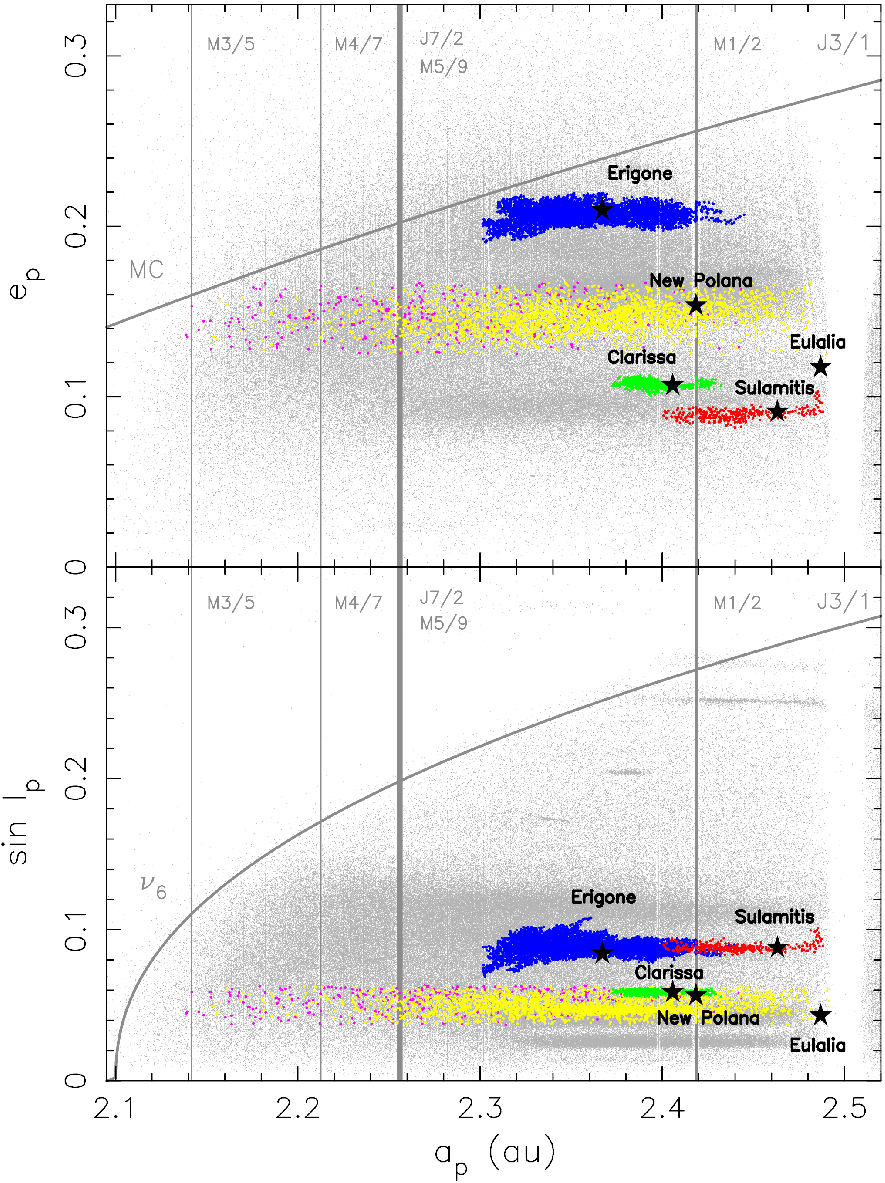} 
 \end{center} 
 \caption{Population context of asteroid (52246)~Donaldjohanson, a member of the C-type Erigone family (blue). Intriguingly, several other C-type families are also located in this region of the main belt, including New Polana and Eulalia, the likely source of spacecraft-visited asteroids (101955)~Bennu and (162173)~Ryugu (see text). These families are highlighted (in violet and yellow, respectively), while the remaining population of asteroids (mostly S-types) is shown using the light-gray symbols. The abscissa is the proper semimajor axis $a_{\rm P}$
  and the ordinate is proper eccentricity $e_{\rm P}$ (top panel) and proper sine of inclination $\sin I_{\rm P}$ (bottom panel). Several orbital resonances are indicated (secular resonance $\nu_6$, Jupiter interior mean motion resonances J7/2 and J3/1, and Mars exterior mean motion resonances M1/2 and M5/9). Stars indicate the largest member of each family.}
 \label{fig1s}
\end{figure}

(52246) Donaldjohanson (hereafter DJ for short) is a primitive inner main belt asteroid (spectrally C-type class), with an estimated average diameter ranging from $3$ to $5$ km \citep[e.g.,][]{mas2011}. DJ is thought to be a member of the Erigone collisional family, which has been estimated to have formed between about $130$ and $270$~Myr ago \citep[e.g.,][]{vok2006a,bot2015,spo2015,mil2019}. The Erigone family is located in the inner part of the main belt, close to several other collisional families of similar spectral taxonomy (Fig.~\ref{fig1s}). These include New Polana and Eulalia families, which are believed to be the respective source families for primitive asteroids (101955)~Bennu and (162173)~Ryugu, visited by the NASA OSIRIS-REx and JAXA Hayabusa2 space missions \citep[see, e.g.,][]{bot2015,tak2024,tat2021}. Unlike New Polana and Eulalia, both crossed by the powerful J3/1 mean motion resonance with Jupiter, the Erigone family is located in a less favorable position to deliver near-Earth objects and meteorites to Earth. This raises the intriguing scientific question as to whether DJ is structurally and compositionally similar to Bennu/Ryugu (and any known primitive meteorite group), or whether it has distinct properties. 

Furthermore, DJ itself appears to be a peculiar object. Ground-based observations reveal
a large light-curve amplitude of $\simeq 1$ magnitude, and a rather long rotation period of $\simeq 252$~hr \citep[e.g.,][]{fer2021}. A possible interpretation is that DJ is quite elongated ($a/c$ body axis ratio $\sim 3$), and that it is a slow rotator, possibly due to thermal torques that have decelerated its spin rate over time \citep[e.g.,][and Sec.~\ref{djeri}]{vok2007}. Both of these characteristics are very distinct from Bennu and Ryugu \citep[$a/c \simeq 1.1$ for both objects and periods respectively of $\simeq 4.3$ and $\simeq 7.6$~hr; e.g.,][]{rob2021}.

A possible explanation for these bulk differences has to do with a different formation mechanism. For instance, it has been suggested that the Erigone family was formed by a sizable cratering event
\citep[e.g.,][]{mil2019}, while both the New Polana and Eulalia families were produced by catastrophic disruption events \citep[e.g.,][]{bot2015}. Depending on the circumstances, this difference might suggest that DJ is a more competent body (perhaps retaining some internal strength) rather than being a rubble pile such as Bennu and Ryugu. Another possibility is that DJ is a less collisionally evolved object based on its relatively young family age and larger size (less than $\sim 270$~Myr and $\sim 4$~km size), while the much older ages of the New Polana ($\sim 1400$~Myr) and Eulalia ($\sim 850$~Myr) families could have resulted in subsequent collisional evolution for Bennu and Ryugu (about 0.49 and 0.90 km in diameter, respectively). Finally, it is possible that there are compositional differences that could affect the formation and bulk properties of these objects. Regarding the latter, a recently published VIS-NIR ($0.5-2.5\;\mu$m) spectra of DJ has shown possible distinctive characteristics compared to Bennu and Ryugu, but the poor signal-to-noise ratio does not allow for any definitive conclusions \citep[see][]{har2024}.

Some of these open issues can be tested by the NASA Lucy spacecraft, which will encounter DJ on April~20, 2025 at a close approach distance of about $900$~km and a relative velocity of $13.4$ km~s$^{-1}$ \citep{lev2021}. The flyby will provide an unique opportunity to image DJ at a highest resolution of about $10$ m~pixel$^{-1}$. The acquired imaging data will be utilized to study DJ's morphology and crater population. We anticipate the quality of the data will be comparable with what was obtained by Lucy during the recent flyby of main belt asteroid (152830) Dinkinesh \citep{lev2024}. The anticipated cratering data will be used to constrain DJ's collisional history, and possibly provide an independent constraint on the formation age of the Erigone family.

In this paper, we use up-to-date observations and modeling efforts to characterize the Erigone family and revise the collisional evolution of DJ in relation to its membership in the Erigone family. We also discuss predictions for the Lucy mission.

\section{Erigone family: home of DJ} \label{family}
Analysis of asteroid families, defined as clusters of asteroids on similar 
orbits as a consequence of high-energy collisions in the main belt, resembles in many
respects a journey of discovery to a new land at the dawn of the exploration era. 
The solitary expedition of this analogy took place more than century ago \citep[see][]{h1918},
when the largest asteroid families were discovered, namely Themis, Eos,
Koronis, and eventually Flora. 

Later, when explorers got closer to the new land and their tools improved, additional details were collected and put on maps. In asteroid families research, this phase effectively started
in the 1970's and 1980's, when our catalogs of asteroids had substantially grown and
mathematical tools needed to define the long-term stable orbital elements of asteroids (the proper
elements) had improved. 

Thanks to these improvements, the core of what is known today as the Erigone family was discovered by \citet{w1979}. At that time, it was simply called family 166 \citep[see also][]{wh1987,w1992}. The cluster only contained a handful of objects and was barely detectable. The discoverer J.G.~Williams made an interesting observation, though, that this family resides very close to Mars-crossing line and could potentially be source of material leaking into the terrestrial planet region. 


The Erigone family is also an interesting case study in how a rapid increase in the number of known asteroids can lead to new challenges. By 2010's, the number of asteroids with reliably determined proper elements surpassed $10^{5}$, making the limited orbital volume of main belt space rather crowded.  As a consequence, the search for families became an increasingly complicated task. \citet{mil2014} attempted to overcome these challenges using an automated objective scheme of family determination. While the Erigone family is listed in their efforts, these authors do not pay particular attention to it. A more detailed analysis of the Erigone family can be found in \citet{spo2015}, \citet{pao2019}, and \citet{mil2019}. At first, these authors proposed a split of the Erigone zone into two clusters, called ``proper'' Erigone and Martes, but they eventually decided that they indeed were a single family\footnote{Curiously, asteroid (5026)~Martes, member of the Erigone family, later
 found its way to family glory by becoming the parent asteroid of an extremely young cluster \citep[see][]{vok2024}.}. The two extremes of the Erigone family in semimajor axis are in fact produced by the Yarkovsky and Yarkovsky-O'Keefe-Radzievski-Paddack (YORP) dynamical evolution of asteroid family members, as described by \citet{vok2006a}. 

In addition, the Erigone family was shown to have primitive surface materials with a low-degree of thermal processing \citep[see][for reviews]{cel2002,metal2015}. Broad-band photometry, as well as the detailed spectroscopy of its largest members, indicate the family belongs to the taxonomic C-type class, while infrared observations show the Erigone members are low-albedo objects (geometric albedo $p_V<0.125$). Taken together, the Erigone family belongs to an interesting group of clusters in the inner main belt whose parent bodies were, at all likelihood, implanted in this zone (see review in Morbidelli et al. 2015). This region is otherwise dominated by S-type asteroids, akin to ordinary chondrites.

A fundamental aspect of asteroid family studies is how to constrain their formation age. Methods that can help on this front included collisional evolution studies, dynamical considerations and, exceptionally, crater counts. In this paper, we will focus on the second method, with obvious implications related to the third, as the Lucy mission will return detailed information about DJ surface. 

Dynamical methods of family age determination, which are based on past orbital (and sometimes even rotational)
evolution of family members, come in different flavors. For very young families, direct
backward orbital propagation of individual orbits allow the user to reconfigure the present-date
family structure to the form resulting directly from the collision event producing the family \citep[see reviews in][]{netal2015,nov2022}. This approach is only applicable for families with ages less than about $20$~Myr. 

For older families, the user must rely on techniques having statistical rather than deterministic natures. A popular method is based on identifying and modeling the traces of the
Yarkovsky thermal drift in the family structure. The Yarkovsky effect causes asteroid smaller than roughly 30 km in diameter to undergo secular changes in semimajor axis over time \citep[see a review in][]{vetal2015}. The extent of this process, if analyzed for orbits of members with different size, may constrain the family age. A challenging part of this analysis, however, is to discern the a priori unknown semimajor axis distribution resulting from the initial ejection of fragments with various sizes. Here, the YORP effect --a rotational alter ego of the Yarkovsky effect-- helps to de-correlate
the initial ejection field from the orbital evolution. 

Details of the method have been developed
in a series of papers by \citet{vok2006a,vok2006b,vok2006c}, with many later variants reviewed in \citet{netal2015}. Importantly, \citet{vok2006a} found the Erigone family to be an ideal test case for the approach and inferred an age of $280\pm 100$~Myr. Subsequent analyses of the Erigone family age with similar or somewhat simplified methods used updated family populations
(as the number of detected asteroids in the family steadily increased), with all reaching similar solutions
\citep[see][]{spo2015,bot2015,car2016,bol2018,pao2019}. The Erigone family was found to be relatively young, with maximum age of a few hundreds of Myr. 

For sake of completeness, we also note the work of \citet{mar1999},
who attempted to determine the Erigone family age based on collisional modeling methods. This study was, however, inconclusive. An updated variant of the same technique by \citet{bro2024} has led to an age of $500\pm 100$~Myr. This age is nearly a factor of two older than those obtained from Yarkovsky-YORP orbital solutions, likely due to the uncertainty of their collisional model and its unknown initial conditions.  We will make our own collisional modeling calculations for Erigone later in the paper.

In what follows, we take a fresh look at the Erigone family just before Lucy's close
encounter with DJ. We will use the most up-to-date catalog of asteroid proper
elements provided in \citet{nes2024}, which allows us to investigate Erigone structure to smaller sizes than before. Our goal is also to provide further justification that DJ is a member of the Erigone family, though a definitive proof is impossible due to the presence of interloping asteroids. Keeping in mind that Lucy's observations will allow us to interpret the nature and history of DJ's surface, we will make an effort to determine a realistic (as opposed to a formal) range of age solutions for the Erigone family.
\begin{figure*}[t!]
 \begin{center}
 \includegraphics[width=0.95\textwidth]{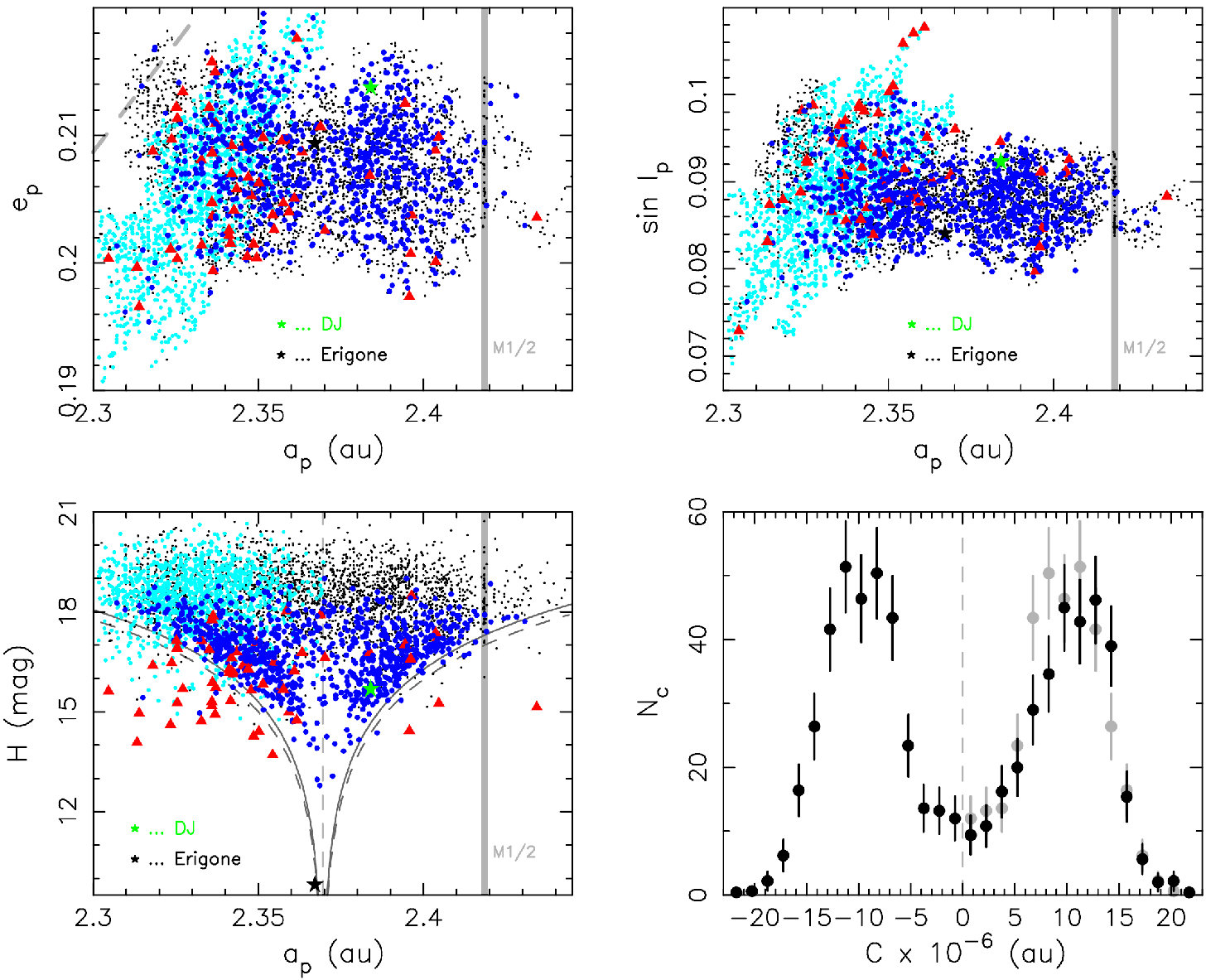} 
 \end{center} 
 \caption{Top panels: Erigone family projected onto 2D planes in proper
 element space: (i) semimajor axis $a_{\rm P}$ vs eccentricity $e_{\rm P}$
 (top left), and (ii) semimajor axis $a_{\rm P}$ vs sine of inclination
 $\sin I_{\rm P}$ (top right). Black dot symbols for all 4925 members identified
 at the HCM cutoff velocity $32.5$ m~s$^{-1}$. Color-coded symbols highlight
 special subgroups of the whole sample: blue symbols circles are 834 low albedo members
 ($p_V <0.125$), red triangles are 55 high albedo members ($p_V>0.125$), and cyan circles are members in the $z_2$ secular resonance (for
 which $|z_2|<0.3$ arcsec~yr$^{-1}$). Location of Erigone and DJ are
 highlighted with black and green stars. The vertical gray line shows location
 of exterior mean motion resonance 1/2 with Mars (M1/2), and the dashed gray curve
 at the top and left panel locates orbits with perihelion $q=1.82$~au (approximate limit
 where where the population becomes efficiently depleted by Mars encounters).
 Bottom and left panel: Erigone family members projected onto the plane defined
 by the proper semimajor axis $a_{\rm P}$ (abscissa) and absolute magnitude $H$
 (ordinate). The same color-coding as in the upper panels. The vertical dashed
 line shows center of the family ($a_{\rm c}=2.3695$~au), and the two solid gray curves
 show the limiting $|C_\star|=1.7\times 10^{-5}$~au lines defined by maximum conjoint
 contrast $r(C_\star,a_{\rm c};\Delta C)$ (see Eqs.~\ref{cline} and \ref{rcrit}), while the
 dashed gray curves correspond to $|C_{\rm fam}|=1.95\times 10^{-5}$~au, the limit used
 for Yarkovsky/YORP chronology modeling. Formal
 members with $H$ smaller than $H_\star$ at the critical lines, for a given $a_{\rm P}$
 value, are deemed unrelated interlopers in the family. Bottom right: Distribution of
 $H\leq 17.5$ magnitude Erigone members $dN(C)$ with $|C|\leq C_{\rm fam}$ values binned
 with $\Delta C= 1.5\times 10^{-6}$~au intervals; only WISE-identified dark
 asteroids used in this dataset. The black symbols, with formally
 adopted $\sigma(C)=\sqrt{dN(C)}$ uncertainty, are are raw data. The gray symbols at
 $C\geq 0$ values mirror the $C\leq 0$ distribution.}
 \label{fig1}
\end{figure*}

\subsection{A fresh look at the Erigone family} \label{family_now}

\noindent{\it Family identification and morphology in the proper element space. }Following closely
the method developed by \citet{km2000,km2003}, \citet{nes2024} determined
synthetic proper elements for more than a million asteroids in the main belt and made
them available through NASA's PDS node%
\footnote{See also \url{https://www.boulder.swri.edu/~davidn/Proper24} and
\url{https://asteroids.on.br/appeal/}.}
(\url{https://pds.nasa.gov/}). We applied the
Hierarchical Clustering Method (HCM) to this database to search for the Erigone family. 

At this stage, we used the traditional variant of the HCM \citep[e.g.,][]{zap1990,netal2015}
operating in 3D space of proper semimajor axis $a_{\rm P}$, proper eccentricity $e_{\rm P}$, and proper sine
of inclination $\sin I_{\rm P}$.  We further verified the identified family with color and
albedo information at later steps. While the background over-density of asteroids in the
Erigone zone is not a critical problem, some degree of experimentation is
needed to fine tune the HCM velocity cutoff $v_{\rm cut}$. The nominal family realization we
shall use here assumes $v_{\rm cut}=32.5$ m~s$^{-1}$. Smaller values of $v_{\rm cut}$ allow us to
identify the family's core but more distant small members in a halo surrounding the core will be missed. Coversely, large values of $v_{\rm cut}$ associate too many unrelated objects to Erigone. For
instance, changing $v_{\rm cut}$ between $28$ and $34$ m~s$^{-1}$ causes the number of members of the
cluster to double from 2777 to 5576. After some trial and error, our nominal Erigone family has 4925 members, including DJ.

The two top panels of Fig.~\ref{fig1} show the projection of the nominal Erigone family onto 2D
planes of $(a_{\rm P},e_{\rm P})$ and $(a_{\rm P},\sin I_{\rm P})$. The bottom left panel shows
the proper semimajor axis at the abscissa and the absolute magnitude $H$ on the ordinate.
The largest member --Erigone-- is approximately at the center of the family. Adopting the
reference central value of the semimajor axis $a_{\rm c}=2.3695$~au (see below), the $\simeq 2.3
\times 10^{-3}$~au distance of Erigone translates to a transverse velocity kick of $\simeq
10$ m~s$^{-1}$ using Gauss equations. This is comfortably smaller than the estimated escape velocity
from this asteroid ($\simeq 40$ m~s$^{-1}$). Family asymmetries at this level may be expected if
the family was produced from a cratering event on Erigone itself. 

The family members extend to both
larger and smaller values of $a_{\rm P}$, with the smallest members reaching the largest distance
from $a_{\rm c}$. No strongly chaotic mean motion resonance crosses the Erigone family zone. A few smaller resonances interact with the family, such as the exterior M1/2 resonance with Mars at $\simeq 2.418$~au, the interior J10/3 resonance with Jupiter at $\simeq 2.331$~au, and the three body resonance (J4,-S2,-1) with Jupiter and Saturn at $\simeq 2.398$~au. They potentially cause some degree of low level depletion \citep[see][]{nm1998,mn1999,gal2011}.

Erigone family belongs to a class of main belt clusters crossed by high-order secular
resonances \citep[see][for review]{car2018}, namely $z_2=2(g-g_6)+s-s_6$ in this particular
case \citep[see][for definition and nomenclature]{mk1992,mk1994}. Erigone members located
in $z_2$, shown by cyan symbols in Fig.~\ref{fig1}, stretch diagonally in the $(a_{\rm P},e_{\rm P})$
and $(a_{\rm P},\sin I_{\rm P})$ plots. The strength of $z_2$ resonance is too small to
cause orbital instability, yet its presence in the family zone may have some interesting implications.
Asteroids whose semimajor axis is affected by Yarkovsky thermal drift may be captured by $z_2$ and thereafter follow this resonance \cite[see][for an example of this phenomenon]{vb2002}. We may thus expect some contamination of the Erigone family by interloping objects in the $z_2$ resonance location. 

Finally, due to moderately large $e_{\rm P}$ values, a potentially destabilizing factor of the Erigone family arises from its location near Mars-crossing orbits at the lowest $a_{\rm P}$ end (shown approximately by the
dashed gray curve on the top and left panel of Fig.~\ref{fig1}). A fraction of the Erigone population
with $H\geq 18$ might have leaked (and is also presently leaking) to the terrestrial-planet
zone (given their low albedo values, these are $\lesssim 1.4$~km dark asteroids). As also shown
on the top and left panel of Fig.~\ref{fig1}, $z_2$ resonance captures may help decrease
this flux into the Mars-crossing region.
\smallskip

\noindent{\it Erigone family in the semimajor axis vs absolute magnitude projection. }Erigone 
family members are distributed in the $(a_{\rm P},H)$ plane in a pattern characteristic
of other families: the largest objects are located near the center, smaller members with diameter $D$ are dispersed
from the center up a distance roughly proportional to $D^{-1}$. At first glance, this may look like the signature of the fragments' ejection velocity field at the moment of family formation. A closer
analysis, however, reveals that this contribution must only be a small fraction of the total. The reasons are as follows.

First, the equivalent velocities required to explain the observed family extension are far larger than the
escape velocity from the parent asteroid \citep[e.g.,][]{cel2004}.  Second, the distribution of
$H\leq 18$ members in the $(a_{\rm P},H)$ plane avoid the center of the family, and instead exhibit a peculiar polarization towards the extreme largest
and smallest $a_{\rm P}$ values. This latter property is not compatible
with any reasonable ejection velocity field.  It only makes sense as a consequence of long-term orbital
evolution driven by the synergy of the Yarkovsky and YORP effects \citep[the latter assisting the former by
tilting spin axes to the direction normal to the orbital plane; see][]{vok2006a}. Erigone is in fact an exemplary case that is well suited for age dating via our Yarkovsky/YORP chronology model.  We shall apply it to the new nominal realization in what follows. 

It is worth mentioning that this polarization pattern
disappears for $H>18$ members. This observation does not contradict the Yarkovsky/YORP model, rather it
is an expected prediction. As discussed by \citet{bot2015}, these small members experience such fast
YORP evolution that, depending on the family's age, they undergo a large number of so-called ``YORP cycles". A YORP cycle is defined as a case where an asteroid goes from a generic initial spin vector state to an asymptotic YORP endstate. Examples of the latter include the body spinning so fast that its sheds mass (and changes shape) or so slowly that it enters into a tumbling rotation state. When the asteroid emerges from this YORP endstate, the pattern begins again, with the body once again spinning up or down toward a YORP endstate.  As
a result, asteroids that undergo many YORP cycles do not continue their steady push toward the most extreme values of $a_{\rm P}$, but instead evolve via a random walk. This causes small family members to take a characteristic Gaussian-type distribution of proper semimajor axis. We note that this concept was developed more fully by \citet{pk2016,pao2019}, and that these papers also include a discussion of the Erigone family.

Accordingly, our Yarkovsky/YORP model of moderately young families, including Erigone, must account for the spin vector evolution of their members. Transport of small members toward extreme-large or extreme-small values of $a_{\rm P}$ by the Yarkovsky effect requires prograde or retrograde rotation states, respectively. This prediction has recently been tested in the case of several families by \citet{dh2023}, with the results matching expectations. In the case of the Erigone family, the available spin-state solutions are unfortunately somewhat limited. We review the current situation in the Appendix~\ref{apper4}.
\smallskip

\noindent{\it Albedo data and $C$-foliation of the $(a_{\rm P},H)$ space. }Further justification of the identified nominal Erigone family arises from the available observations at infrared wavelengths, as well as visible multi-color photometry \citep[see][]{par2008,mas2011,mas2013}. Here we use data obtained by the Wide-field Infrared Survey Explorer (WISE) \citep{mas2011} that provided diameter and albedo values for more than 125,000 asteroids. \citet{metal2015} used the 2015 edition of the PDS families identification \citep{netal2015} and identified 716 Erigone members with WISE data. They determined the predominance of dark-albedo objects, with a median albedo of $0.05\pm 0.01$ \citep[this is in agreement with color-indexes obtained using observations of the Sloan Digital Sky Survey (SDSS) being compatible with predominant C-complex taxonomy; e.g.,][]{par2008,metal2015,mor2016,har2024}. 

\begin{figure}[t!]
 \begin{center}
 \includegraphics[width=0.47\textwidth]{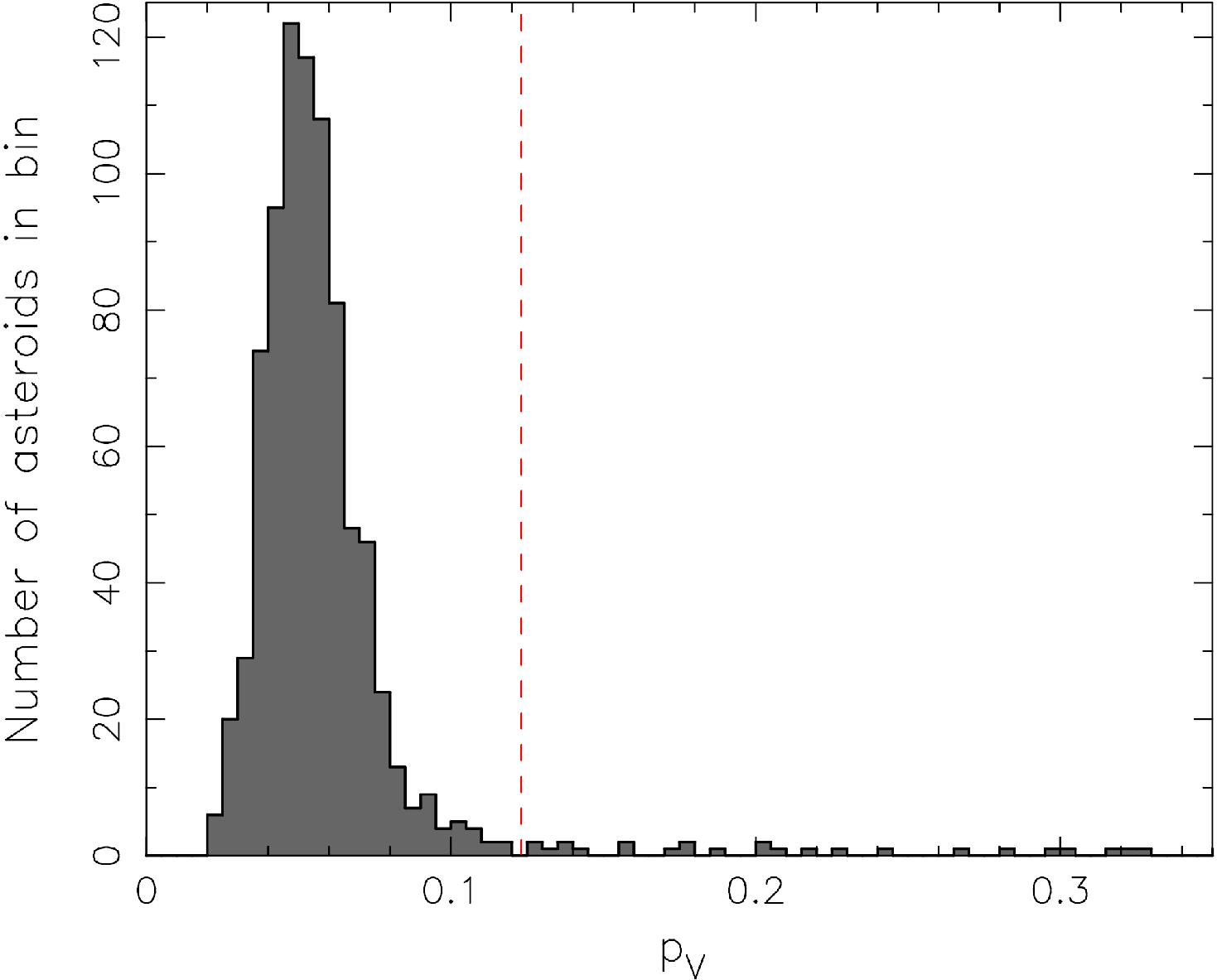} 
 \end{center} 
 \caption{Distribution of the geometric albedo values $p_V$ for 862 members in the Erigone family delimited by the dashed gray curves in the $(a_{\rm P},H)$ shown in Fig.~\ref{fig1} (the ordinate is a number of asteroids in $0.005$ wide albedo bins). The bulk of the family asteroids have dark albedo values with $p_V\leq p^\star_V=0.125$ (red dashed line) confined in a narrow peak about the median ${\bar p_V}=0.054$, followed by a tail sub-population with $p_V$ values terminated by $p^\star_V$ \citep[see also][]{mas2013}. No clustering of $p_V$ values is seen above $p^\star_V$ among a group 34 bright objects, suggesting they are interlopers in the family.}
 \label{figa}
\end{figure}

Our nominal Erigone family contains 889 members for which
WISE provides size and albedo values. The albedo distribution of members of the Erigone family shows a high-albedo tail that becomes discontinuous for albedos grater than 0.125.  Figure~\ref{figa} shows the distribution of the geometric albedo values $p_V$ of this sample. The albedo limit at $p_V^\star$ (shown by the red dashed line) clearly terminates the bulk of the dark population. Thus, adopting the albedo value of $p_V^\star= 0.125$ as a criterion to divide the sample to dark objects (defined here as low albedo) and bright objects (defined here as high albedo), we find 834 dark members and 55 bright members (the WISE sample in the family is highlighted by blue dots --dark members-- and red triangles --bright
members-- in Fig.~\ref{fig1}). The median albedo of the dark sample is ${\bar p_V}=0.054$, very close to the results of \citet{mas2013,metal2015}. 

Overall, the bright interlopers only represent $6.1$\% of the nominal family. Restricting ourselves to the $(a_{\rm P},H)$ region within the two dashed gray lines shown in the bottom and left panel of Fig.~\ref{fig1}, we find that the population of dark members is 828 and the sample of bright interlopers drops to 34 (representing now only $3.9$\%).
Likewise, the region outside the two dashed gray lines contains only 6 dark objects but 17 bright
ones. Furthermore, these ``exterior'' bright interlopers mostly have $a_{\rm P}\leq 2.35$~au; they show an affinity to the orbital location of the $z_2$ secular resonance. This supports our suspicion
that a significant fraction of the low-level contamination of the Erigone family may be associated
with interlopers migrating along this resonance (possibly over Gyrs).  These objects would therefore sample 
various parts in the inner
main belt. This justifies our decision to restrict our further analysis to a sub-sample of the nominal Erigone
family delimited in the $(a_{\rm P},H)$ plane by the interior of the dashed gray lines. This approach has been adopted by many previous studies and was reviewed in Sec.~4 of \citet{netal2015}.
In the Appendix~\ref{apper1} we provide a method to quantitatively determine the family formal
center $a_{\rm c}$ and its borderline gray lines shown in Fig.~\ref{fig1}.

In order to further analyze the Erigone family, we need to define our notation and variables for our
Yarkovsky/YORP chronology model. Here we introduce a method of folding data in the 2D $(a_{\rm P},H)$ plane onto a
suitable 1D variable \citep[see][for more details]{vok2006a}. To that end, we define parameter $C$ using
an implicit relation
\begin{equation}
 H(C,a_{\rm P};a_{\rm c})=5\,{\rm log}\left(\frac{a_{\rm P}-a_{\rm c}}{C}\right) \; , \label{cline}
\end{equation}
where $a_{\rm c}$ is a free parameter of symmetry: given a certain value of $H$, positive and
negative values of $C$ correspond to $a_{\rm P}$ values symmetric with respect to $a_{\rm c}$. In the case of Erigone family we adopt $a_{\rm c}=2.3695$~au (see the Appendix~\ref{apper1} for a formal method that justifies this value). 

The relevance of the $C$-parameter for our Yarkovsky model consists of the fact that asteroids 
starting at $a_{\rm c}$ (or very close) and maintaining a constant drift rate of 
$da_{\rm P}/dt$ would, in a given time $T$, reach the same $C$-isoline independently of their size $D$.
The reason is because $da_{\rm P}/dt\propto D^{-1}$. Asteroids drifting at the maximum possible Yarkovsky rate,
namely those having extreme obliquity values of $0^\circ$ or $180^\circ$, would reach a maximum isolines
$|C_{\rm fam}|=(da/dt)_{1329}\,\sqrt{p_V}\,T$ \citep[e.g.,][here $(da/dt)_{1329}$ is a drift rate of
$D=1329$~km large asteroid]{vok2006a,netal2015}. This ``wavefront'' of asteroids is close to the gray lines shown in
bottom and left panel Fig.~\ref{fig1}, which are in fact defined by Eq.~(\ref{cline}) for a certain $C$
values. 
\begin{figure*}[t!]
 \begin{center}
 \includegraphics[width=0.95\textwidth]{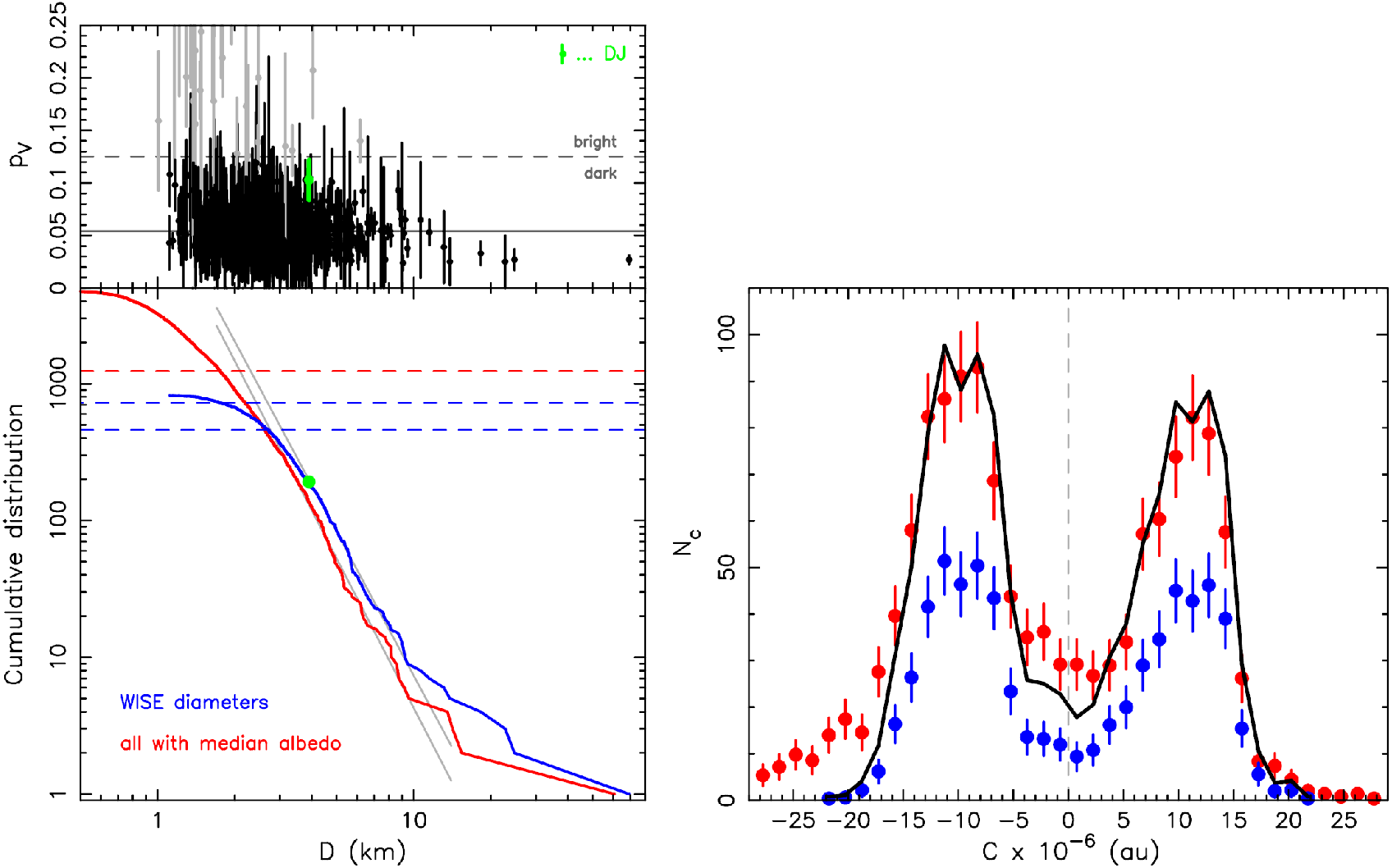} 
 \end{center} 
 \caption{Left and top panel: Geometric albedo values $p_V$, with formal uncertainties,
  determined for 889 Erigone family members by analysis of WISE observations. The sample
  is dominated by 834 dark objects ($p_V< 0.125$) shown by black symbols. The albedo
  values for a bright group (55 objects with $p_V> 0.125$) are shown using gray symbols
  (some would fall even beyond the upper limit $0.25$ of the plot). DJ's value
  is shown in green. The horizontal gray line at ${\bar p_V}=0.054$ indicates the median
  value of the dark sample. Note the largest objects ($D>10$~km) have systematically
  $p_V< {\bar p_V}$. Left and bottom: Cumulative size distribution of Erigone family
  members: red curve for all 4925 members assigning to all median WISE albedo ${\bar p_V}=
  0.054$ for the dark sample and adopting their absolute magnitudes, blue curve is
  the distribution of size for 834 members with WISE dark albedo. DJ location
  marked by green symbol. The grey lines are approximations with power-law distribution
  $N(>D)\propto D^{-\alpha}$ in the $4$ to $10$~km range with $\alpha\simeq 3.63$ for
  the whole population (red) and $\alpha\simeq 3.50$ for the WISE sub-sample (blue). The
  horizontal dashed lines correspond to $H=17.5$ magnitude limit. This is unique for the
  whole sample in red (where sizes are computed from magnitudes using fixed albedo
  value), while $H=17.5$ asteroids are in a certain range in the WISE sample, because
  individual objects have slightly different $p_V$ values. Right panel: Distribution of
  $dN(C)$ binned with $\Delta C= 1.5\times 10^{-6}$~au intervals for $H\leq 17.5$ magnitude
  Erigone members (formal $\sqrt{dN(C)}$ uncertainties shown by vertical bar): the blue symbols
  for the sub-sample of asteroids for which WISE determined $p_V<0.125$ albedo values and
  $|C|\leq C_{\rm fam}$ (the same as black symbols at the bottom right panel on Fig.~\ref{fig1}),
  the red symbols for all Erigone members. The black line is simply the blue distribution
  multiplied by a factor $1.9$.
  A good match to the total population suggests (i) a small fraction of bright interlopers
  in the Erigone family up to the $H=17.5$ magnitude limit, and (ii) there is a $1.9$ incompleteness
  factor of the WISE-observed sample. The mismatch for $C<-20\times 10^{-5}$~au is likely
  due to contamination by bright objects transported along the $z_2$ secular resonance.}
 \label{fig2}
\end{figure*}

With $C$ defined, we can now represent the family population using a distribution ${\cal D}(C)$, 
such that number of family members between $(C,C+dC)$ is $dN(C)={\cal D}(C)\,dC$. The black symbols at the bottom 
and right panel in Fig.~\ref{fig1} show $dN$ values for bins $dC=1.5\times 10^{-6}$~au that are restricted
to Erigone members (i) with dark albedo values by WISE ($p_V\leq 0.125$), and (ii) those having $H\leq 17.5$. 
The first condition should minimize contamination by interlopers, while the latter is given
by our intention to use these data for the Yarkovsky/YORP chronology of the Erigone family. We use the latter because the
$H\leq 17.5$ population shows the optimum telltale signature of Yarkovsky/YORP synergy over the
family age.  The family members pile up to the $a_{\rm P}$ values along the above defined $C_{\rm fam}$
isoline and deplete the center of the family. As discussed above, the population of smaller members with
$H>17.5$ falls into a different Yarkovsky/YORP regime, when $T$ is large compared to the timescale of YORP
cycles and asteroid orbits perform a random walk in $a_{\rm P}$ rather than steady flow. We avoid using this
regime for family chronology, since modeling repeated YORP cycles is challenging given our current state of knowledge. 

The $dN(C)$ distribution
shown in Fig.~\ref{fig1} shows distinct maxima at $C_{\rm max}\simeq \pm 1.1\times 10^{-5}$~au and a minimum in the
center. This reflects the concentrations of the Erigone members at the extreme $a_{\rm P}$ values for a given $H$
in the $(a_{\rm P},H)$ plane. For an optimum choice of $a_{\rm c}$, the distribution $dN(C)$ should be nearly symmetrical in $C$. This claim can be visually tested by the gray symbols, which just flip the distribution with negative
$C$ to their symmetric positive $C$ values. While not completely symmetric, the degree of symmetry is deemed
satisfactory for our work (the observed difference may be due a slight anisotropy of the initial ejection
velocity field or interloper contamination in the $a_{\rm P}\leq 2.35$~au). Given that we do not intend to model
such details, we shall use a fictitious distribution 
\begin{equation}
 dN_{\rm sym}(C)=\frac{1}{2} \left[dN\left(C\right)+dN\left(-C\right)\right] \label{dnsym}
\end{equation}
with an enforced symmetry for our Yarkovsky/YORP chronology model.
\smallskip

\noindent{\it Size distribution of Erigone family members. }The left panels of Fig.~\ref{fig2} show albedo and size data for Erigone family 889 members determined using WISE observations \citep{mas2011,mas2013}: (i) albedo values with their uncertainty (top part), and (ii) cumulative size distribution (bottom part). In case of multiple inputs for a given body, we first drop data with too few observations in the W3 passband (namely less than 6), and take the average albedo value from the remaining set.
Interestingly, the largest members with $D>10$~km have albedo values
smaller than the median ${\bar p_V}=0.054$ of the whole sample of dark objects. This could suggest a slight
trend toward higher albedos for small members.  More likely, there is some unrecognized systematic component in
the uncertainty for these small asteroids. In fact, the case of DJ with $p_V=0.103\pm 0.019$ may
be an exemplary case, as we argue below. 

As for the size
distribution shown at the bottom part of the plot, we show the size-frequency distributions of 834 dark objects provided
by WISE observations (blue curve) and those for 4925 members in the nominal family.  For the latter, we assign to all objects the median albedo value $0.054$ (and using their Minor Planet Center absolute
magnitude values; red curve). The difference between the two size-frequency distributions (SFD) is small but noticeable. It arises from the lower albedo values of Erigone's family largest members. This mismatch illustrates the uncertainty in other asteroid family studies
that infer the family's size frequency distribution from the absolute magnitude distribution data while also assuming the family albedos correspond to the taxonomic class of the largest members. 

Finally, the right panel in Fig.~\ref{fig2} shows the $dN(C)$ distribution for Erigone family members with $H\leq 17.5$ (in the Appendix~\ref{apper2} we argue this sample in complete). For reference, we repeat the distribution for
the 640 dark members with $|C|\leq C_{\rm fam}$ from Fig.~\ref{fig1} (blue symbols), while the red
symbols show the whole Erigone sample. The two distributions
are similar to each other. Multiplying the former by a factor $1.9$ allows one to obtain an excellent match to
the latter (this is shown by the black line). This confirms that there are few interloping objects in either sample of Erigone family members. In fact, the factor $1.9$ is a good
estimate of incompleteness for WISE asteroid observations with $H\leq 17.5$ in the Erigone
zone. This incompleteness does not stem from photometric sensitivity of WISE, as the fluxes in
its W3 passband should correspond to the IR equivalent magnitude $\lesssim 10.25$ \citep[the limit of the
instrument; e.g.,][]{wise2011b}.  Instead, it stems from some Erigone members that did not happen to 
geometrically fit in the WISE field of view. The growing mismatch beyond the family limit in the
$C<-20\times 10^{-6}$~au bins is likely an expression of the interloper population located in the
$z_2$ secular resonance (see Fig.~\ref{fig1}).
\begin{figure*}[t!]
 \begin{center}
 \includegraphics[width=0.95\textwidth]{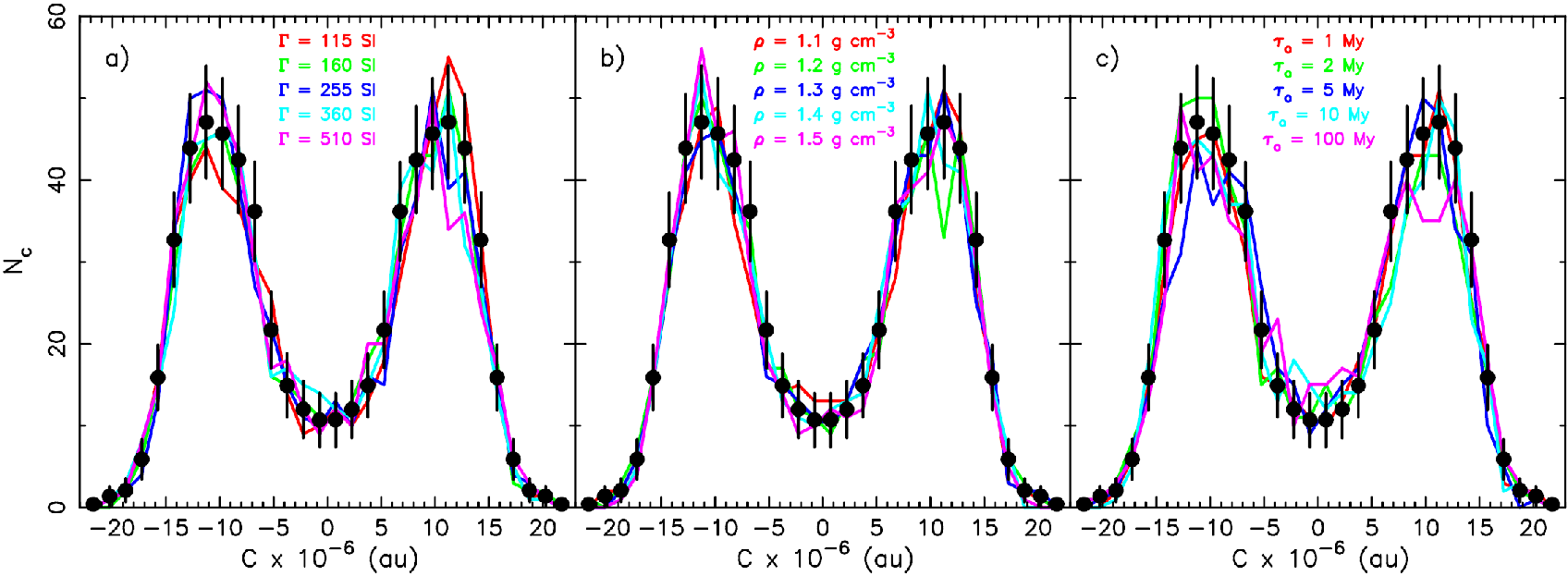} 
 \end{center} 
 \caption{Best-fit solutions from a set of trial simulations sampling each time only one of the
  parameters ${\bf p}_2=(\rho,\Gamma,\tau_0)$ in our model: (i) fixed values $\rho=1.3$ g~cm$^{-3}$
  and $\tau_0=1$~Myr with five values of $\Gamma$ listed in labels and color (left panel), (ii) fixed
  values $\Gamma=160$ SI and $\tau_0=1$~Myr with five values of $\rho$ listed in labels and color
  (middle panel), and (iii) fixed values $\rho=1.3$ g~cm$^{-3}$ and $\Gamma=160$ SI with five values of 
  $\tau_0=1$~Myr listed in labels and color (right panel). The black symbols with uncertainty intervals
  show the data, namely distribution $dN_{\rm sym}(C)$ (Eq.~\ref{dnsym}), with $|C|\leq C_{\rm fam}$ values binned
  using $\Delta C= 1.5\times 10^{-6}$~au intervals for $H\leq 17.5$ magnitude Erigone members with
  dark WISE albedo values. In each case the solution is statistically acceptable, indicating correlations
  between the parameters. In the simulations (i) and (ii), left and middle panels, the Erigone age $T$
  approximately scales with the tested parameter, i.e., $T\propto \Gamma$ and $T\propto \rho$ as 
  expected from analytic formulas for $da/dt$ \citep[e.g.,][]{vetal2015}. In the last simulation (iii),
  right panel, $T$ depends on $\tau_0$ only weakly, but the $\chi^2$ value increases for larger
  $\tau_0$.}
 \label{fig3}
\end{figure*}
\smallskip

\noindent{\it Yarkovsky/YORP chronology revisited. }We now revisit our determination of the Erigone
family age $T$ using the Yarkovsky/YORP method. The general outline of our approach is given in
Appendix~\ref{apper3}. More details can also be found in \citet{vok2006a} and \citet{bot2015}. The goal
is to match the family distribution $dN_{\rm sym}(C)$ in the $C$ parameter using a suitable model
prediction $dM(C;{\bf p})$. The model performance is evaluated using a target function
\begin{equation}
 \chi^2 = \sum_i \left(\frac{dN_{{\rm sym},i}-dM_i({\bf p)}}{\sigma_i}\right)^2 \; , \label{tg}
\end{equation}
where $\sigma_i=\sqrt{dN_{{\rm sym},i}}$ and the summation runs over $|C|\leq C_{\rm fam}$ bins. The formal
confidence boundary of the parameter solution stems from the dimensionality of the ${\bf p}$-space.
With 6 parameters used, our 90\% confidence level corresponds to a hyperspace delimited by
$\delta \chi^2=10.6$ increase of the target function over the best-fit value \citep[e.g.,][]{nr2007}.
\begin{figure}[tp]
 \begin{center}
 \includegraphics[width=0.45\textwidth]{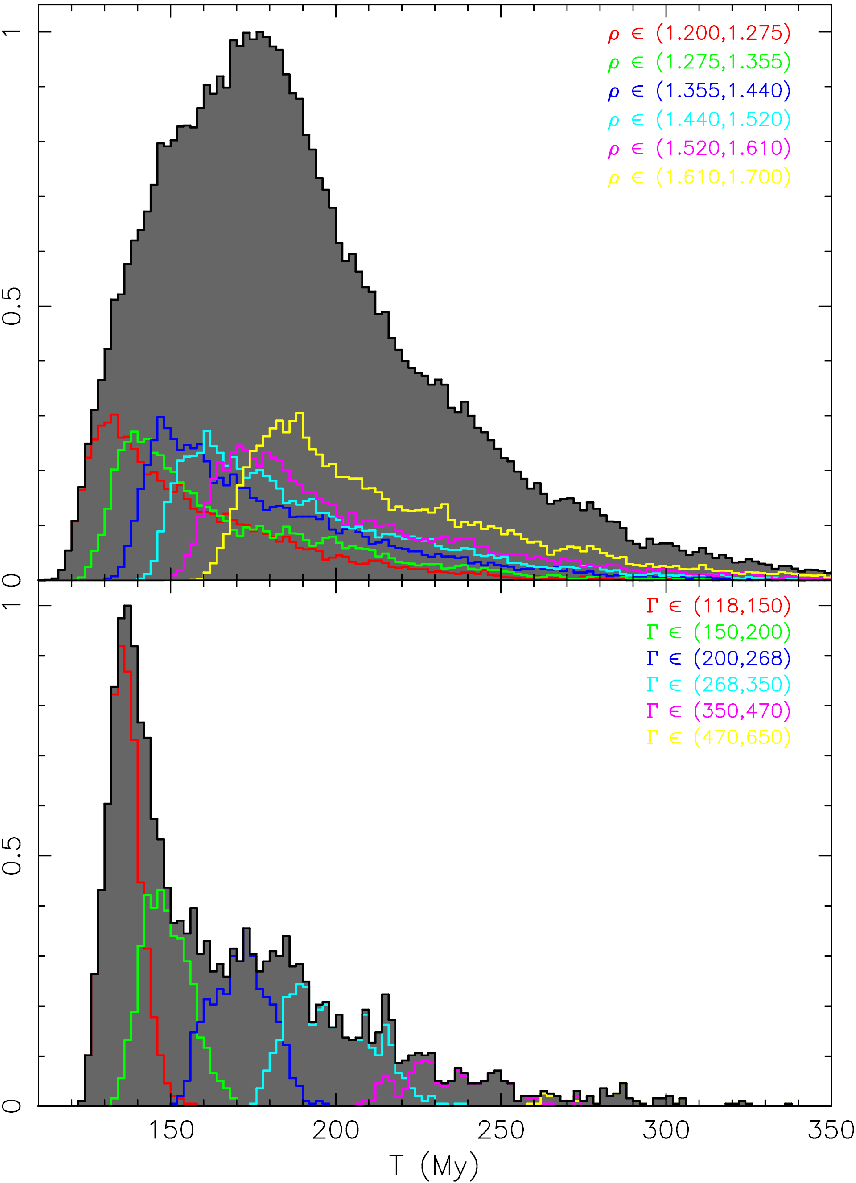} 
 \end{center} 
 \caption{Top panel: Distribution of the Erigone family age $T$ from a set of simulations in which all
  parameters ${\bf p}_2=(\rho,\Gamma,\tau_0)$ uniformly sampled the following interval of values (grey
  histogram with black outline): (i) $\rho \in (1.2,1.7)$ g~cm$^{-3}$, (ii) $\Gamma \in (115,710)$ SI,
  and (iii) $\tau_0 \in (0.5,100)$~Myr (the latter two in log-measure). The median age is $186$~Myr.
  The color-coded distributions are for sub-samples characterized by distinct bulk densities: 
  (i) $\rho \in (1.2,1.275)$ g~cm$^{-3}$ (red), (ii) $\rho \in (1.275,1.355)$ g~cm$^{-3}$ (green), 
  (iii) $\rho \in (1.355,1.44)$ g~cm$^{-3}$ (blue), (iv) $\rho \in (1.44,1.52)$ g~cm$^{-3}$ (cyan), 
  (v) $\rho \in (1.52,1.61)$ g~cm$^{-3}$ (magenta), and (vi) $\rho \in (1.61,1.7)$ g~cm$^{-3}$ (yellow).
  All distributions normalized by maximum of the total distribution. The lower-density
  solutions imply a smaller family age. Bottom panel: Distribution of the Erigone family age $T$
  from a set of simulations in which bulk density was fixed to $1.3$ g~cm$^{-3}$, but the remaining
  two parameters ${\bf p}_2=(1.3,\Gamma,\tau_0)$ sampling the the following interval of values (grey
  histogram with black outline): (i) $\Gamma \in (118,650)$ SI, and (ii) $\tau_0 \in (0.5,100)$~Myr
  (both in log-measure). The color-coded distributions are for sub-samples characterized by distinct
  values of the thermal inertia (all in SI units): (i) $\Gamma \in (115,150)$ (red), (ii) $\Gamma
  \in (150,200)$ (green), (iii) $\Gamma \in (200,268)$ (blue), (iv) $\Gamma \in (268,350)$ (cyan), 
  (v) $\Gamma \in (350,470)$ (magenta), and (vi) $\Gamma \in (470,650)$ (yellow). The lower
  inertia solutions imply a smaller family age. As demonstrated by the two panels, the family
  age becomes more constrained if one (or more parameters) are determined.}
 \label{fig4}
\end{figure}
\begin{figure}[t!]
 \begin{center}
 \includegraphics[width=0.47\textwidth]{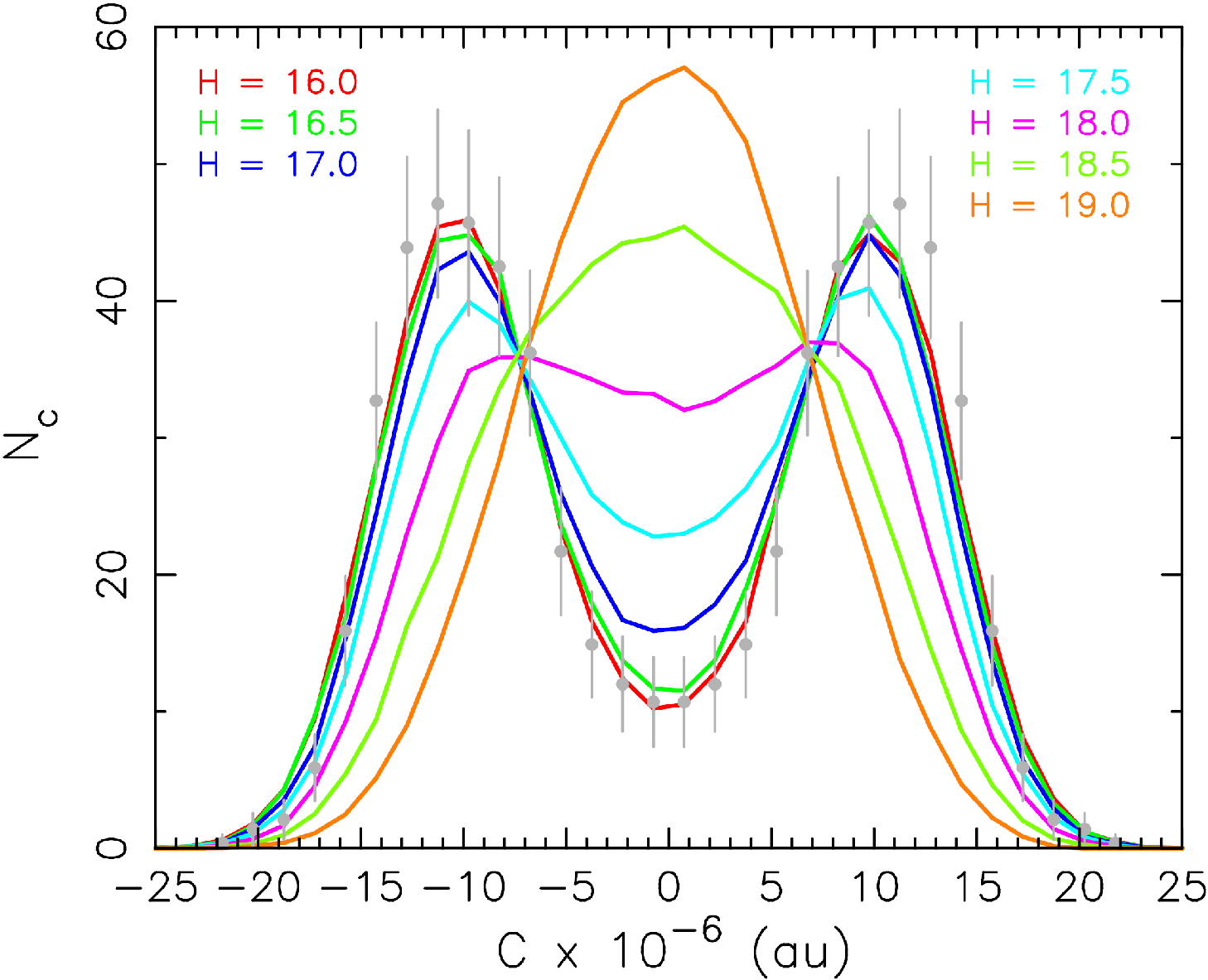} 
 \end{center} 
 \caption{Model prediction of the $C$-parameter distribution $dM(C;{\bf p})$ determined from test
  simulations, in which all asteroids were given the same absolute magnitude value $H$ (color-coded
  with labels) and the parameters ${\bf p}$ were the best-fit set from the Yarkovsky/YORP chronology
  method. The latter used only $H\leq 17.5$ Erigone dark members with their WISE-determined sizes. 
  Here in each magnitude class we assigned sizes from $H$ and $0.054$ albedo value, and pushed the
  tested $H$ values beyond the $17.5$ limit. In the simulations with $H\leq 17.5$ we recover the
  two-hump $dM(C;{\bf p})$ distribution fitting the data (grey symbols), but runs with $H\geq 17.5$
  result in a single-peaked $dM(C;{\bf p})$ distribution due to asteroids random walk in $a_{\rm P}$
  related to quickly decreasing YORP-cycle timescale with size. This transition at approximately
  $1$~kilometer size is indeed seen in the family-members distribution in the $(a_{\rm P},H)$ plane
  (left and bottom panel on Fig.~\ref{fig1}).}
 \label{fig5}
\end{figure}

As for the parameters ${\bf p}$, we split them in two groups: ${\bf p}=({\bf p}_1;{\bf p}_2)=(T,v_5,
c_{\rm YORP};\rho,\Gamma,\tau_0)$. The first set, ${\bf p}_1$, has been considered already by
\citet{vok2006a}. It consists of (i) the family's age $T$, (ii) the initial ejection velocity $v_5$ of $D=5$~km
fragments (assuming the ejection velocity $v_5\,(5\,{\rm km}/D)$ of fragments with an arbitrary diameter $D$),
and (iii) a scaling factor $c_{\rm YORP}$ of the reference YORP torque taken from \citet{cv2004}. In the production simulations, we sampled the following range of these parameters: $50$ to $500$~Myr for $T$,
$0$ to $50$ m~s$^{-1}$ for $v_5$, and $0$ to $2$ for $c_{\rm YORP}$. 

The second set, ${\bf p}_2$, consists of (i) the bulk density $\rho$ of the Erigone family members, (ii) their surface thermal inertia $\Gamma$, and (iii) the characteristic timescale $\tau_0$ of the YORP
strength modification for $D=1$~km members (assuming its size dependence is $\propto \tau_0\,\sqrt{D/(1\,
{\rm km})}$). The last parameter corresponds to what \citet{bot2015} called variable or stochastic
YORP. This concept arises from the fact that YORP torques have a substantial dependence on the small-scale irregularities of
asteroid's shape. As a result, asteroid shape changes caused by sub-catastrophic impacts or landslides may change the strength of YORP torques for kilometer size asteroids. The role of these parameters on age calculations has yet to be fully evaluated, and in some cases it could be meaningful. 

Here we consider $\tau_0$ in the $0.5$ to $100$~Myr range. Concerning the bulk density and surface
thermal inertia, we will use observed parameters for Bennu and Ryugu because these bodies are thought to be close analogs of what we might expect for Erigone kilometer-size family members.  Bennu's mean bulk 
density was found to be $1.2$ g~cm$^{-3}$ \citep[e.g.,][]{lau2019,sche2020}, while its mean
surface thermal inertia was $300\pm 30$ J~m$^{-2}$~K$^{-1}$~s$^{-1/2}$ \cite[denoted SI units below for
short;][]{roz2020}. Given that specific regions of Bennu, such as its equatorial ridge zones, were found
to be modestly lower in density, slightly higher bulk density values may also be possible. Similarly, specific fine-grained 
locations or anomalous rock population (likely covered with fine dust layer) on Bennu's surface have
thermal inertia values as low as $\simeq 150-200$ SI units. The larger heliocentric distance of Erigone
family members, which means lower mean temperatures, may make it worthwhile to explore these lower thermal inertia values. 

Ryugu studies show it also has a mean bulk density of $1.2$ g~cm$^{-3}$ \citep[e.g.,][]{wat2019} and a surface thermal inertia
close to $300$ SI units \citep[e.g.,][]{oka2020}. We note that some studies report lower
values such as $\Gamma=225\pm 45$ SI units in \citet{shi2020}. On the other hand, possible meteorite analogs of Erigone
family members \citep[CM or CR chondrites; e.g.,][]{bro2024} often have larger densities and thermal inertia values.  
We caution that these objects probably represent a biased sample; weaker objects may be unable to survive passage
through Earth's atmosphere. Still, rather than limit ourselves in our runs, we opted to explore bulk density 
values in the range $1.2$ to $1.7$ g~cm$^{-3}$ and surface thermal inertia in the $100$ to $650$ SI units range. For the sake of simplicity, we do not account for any possible size dependence in thermal inertia for this study \citep[see][for discussion and results]{bol2018}.

Before discussing results from the suite of production simulations, where we sample the entire 6D parameter
space of our model, we first ran a set of trial simulations.  Here we tested the dependence of the results on
each of the ${\bf p}_2$ parameters individually (but taking into account the ${\bf p}_1$ parameters
over their range). Results are shown in Fig.~\ref{fig3}. In each of the trial simulations, we find
statistically acceptable solution. This hints that parameter correlations exits, which in fact is expected.

In the left and middle panels, we confirm the Erigone age scaling $T\propto \rho$ and $T\propto \Gamma$
that follow from the corresponding inverse scaling of the Yarkovsky drift rate $da/dt$ \citep[the latter
due to predominant larger thermal parameter regime; e.g.,][]{vetal2015}. Therefore, younger ages are obtained for 
the Erigone family provided densities and/or thermal inertia values are small (or some combination of the two). Likewise,
older family ages come from larger densities and/or thermal inertia values (or some combination of the two). 

We find the age does not depend strongly on the $\tau_0$ value, but $\tau_0 > 50$~Myr does provide poorer fits
to the observational data. The reason is because there is a significant contrast between the maximum 
of $dN_{\rm sym}(C)$ at $C_{\rm max} \simeq 1.1\times 10^{-5}$~au at the family center $C=0$.  It requires the evolution of the rotation rates to be delayed compared to obliquities via the YORP effect. This behavior is well represented by
the random-walk effect in rotation rate provided by variable YORP \citep[see, e.g., Appendix
in][]{bot2015}.

Figure~\ref{fig4} shows results from our production simulations. We focus here on the solution for the age of the Erigone family. The formal best-fit solution with $\chi^2=6.46$ corresponds 
to the following set of parameters: $\rho=1.5$ g~cm$^{-3}$, $\Gamma=135$ SI, $\tau_0=0.6$~Myr, $T=155$~Myr,
$v_5=24$ m~s$^{-1}$, and $c_{\rm YORP}=1.5$. We caution that because there are multiple correlations, the admissible range of solutions is arguably more important than the best-fit solution. Using the 90\% confidence level limit, we projected its volume in 6D ${\bf p}$-space onto the 1D distribution of the family age $T$. These values are shown in the upper panel of Fig.~\ref{fig4}. The median age ${\bar T}=186$~Myr with an asymmetric
$C$ distribution. The youngest age is $\approx 115$~Myr, while the oldest ages extend to $350$~Myr due to a stretched
tail. 

To understand this behavior, we broke the distribution into six categories in density, as shown by
the color histograms in the upper panel of Fig.~\ref{fig4}. Apart from the obvious $T\propto \rho$ scaling,
there is a self-similarity that takes place as the individual distributions correspond to various density values.
The long tails correspond to simulations with large thermal inertia values ($\Gamma$).
This result is confirmed in the bottom panel of Fig.~\ref{fig4}, where we show the 90\% confidence level
age distribution for a fixed value $\rho=1.3$ g~cm$^{-3}$: the color distributions break the simulations into
different $\Gamma$ values for various intervals. As anticipated, the age solutions larger than $\simeq 250$~Myr
are in the long tail and are obtained for large $\Gamma$ values.

Finally, we ran checks to make sure the parameter set ${\bf p}$ of our solutions does not contradict the Erigone family
structure in the $(a_{\rm P},H)$ plane for $H\geq 18$ (left and bottom panel on Fig.~\ref{fig1}). The
Yarkovsky/YORP model is based on matching the polarization of $H\leq 17.5$ members in $a_{\rm P}$ toward
``extreme'' values defining the $C_\star$ isoline. This behavior is not observed in the Erigone $H\geq 18$ population, so the accepted solutions must satisfy this constraint. 

We ran the Yarkovsky/YORP
model with the best-fitting set of parameters ${\bf p}$ mentioned above. For each time, we took a synthetic
(not real) population of 640 asteroids of a fixed magnitude $H$, sampling values from $16$ to $19$ with
an increment of $0.5$ (their sizes were derived from the standard magnitude-size relation assuming
the median albedo value $0.054$ of the family members). We ran the model for $T=155$~Myr and determined
the final $dM(C;{\bf p})$ distribution for each magnitude class separately. The results are
shown in Fig.~\ref{fig5}. 

The simulations that propagated $H\leq 17.5$ asteroids provided the
double peak distribution needed to match the data (shown in light-gray symbols for completeness). Moving to results in simulations where $H\geq 18$, the resulting distribution changes, having just a
single maximum at the family center. As discussed above, this behavior follows from the asteroids performing
a random walk in $a_{\rm P}$ due to short YORP cycles instead of a steady flow
towards the $C_\star$ limit. While our fits are satisfactory, we caution the reader that proper modeling of this evolutionary regime is difficult with our approximate model. For that reason, we do not use it
in determining the family's age.
\smallskip

\noindent{\it Size of Erigone family parent body. }One of the more challenging aspects of asteroid family studies is estimating the original size of the parent body. The reason is that in a family forming event, a considerable amount of mass is placed into fragments that are smaller than the observation limit in the main belt \citep[e.g.,][]{dur2007}, which is probably near a few kilometers in diameter.  Even if we could somehow magically see all of this material in the present day, we would still have issues because sub-km objects are readily ground down over hundreds of Myr by collisional processes \citep[e.g.,][]{betal2015}.  

One way to get around this problem is to consider the largest members of a family. They are the least affected by collisional evolution and typically have very slow Yarkovsky drift rates. The combination implies the these bodies are probably the most unchanged within a family. Accordingly, these bodies can be compared to the SFDs of results from numerical hydrocode simulations of impacts. Specifically, here we will make comparisons to the 161 simulations conducted to study asteroid satellite formation by \citet{dur2004}. Numerical impact simulations must conserve mass, so if we can find a match between a scaled version of a model family SFD and the observed family SFD for the largest fragments, the results can give us insights into the initial parent body size. Our method is as follows.
\begin{figure}[t!]
 \begin{center}
 \includegraphics[width=0.47\textwidth]{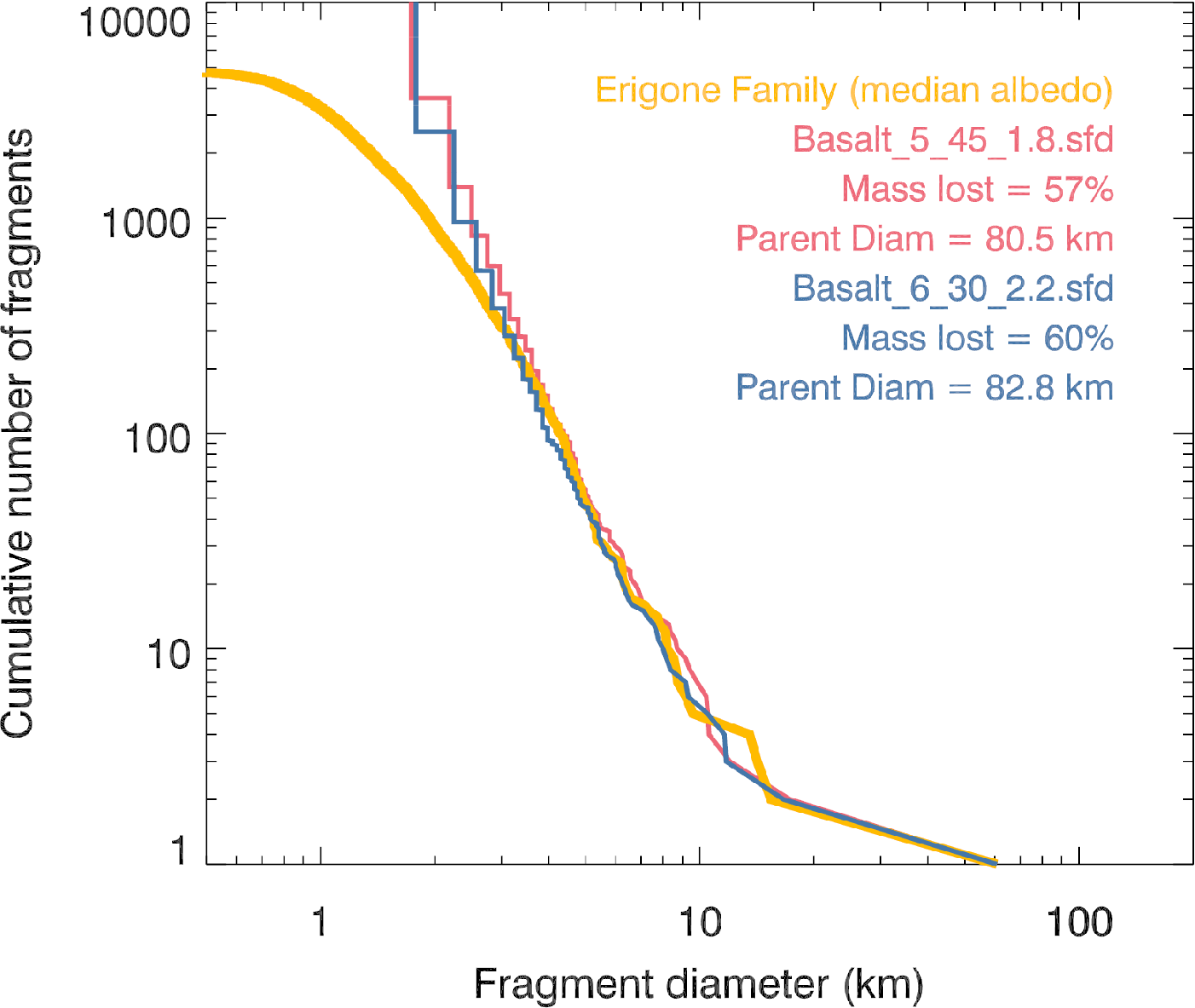} 
 \end{center} 
 \caption{Two modeled fragment SFDs compared to the observed Erigone family SFD. The latter is the gold curve, and it comes from Fig.~\ref{fig2}. The modeled fragment SFDs come from \citet{dur2007}. For the red curve, defined as "Basalt\_5\_45\_1.8", a 100-km-diameter solid basalt target was hit at $5$ km~s$^{-1}$ at a $45^\circ$ impact angle, with the logarithm of the target-to-projectile mass ratio being $1.8$ (i.e., a $25$~km diameter projectile). For the blue curve "Basalt\_6\_30\_2.2", a 100-km-diameter target was hit at $6$ km~s$^{-1}$ at a $30^\circ$ impact angle by a $18$~km diameter projectile.  The fragment SFDs from both runs were scaled to fit the observed Erigone family SFD. For the red curve, the net mass in the fragment SFD, excluding the largest remnant, is 57\% of the parent body, which we estimate was $80.5$~km in diameter, while for the blue curve, the values are 60\% and $82.5$~km, respectively.}
 \label{fig7}
\end{figure}

First, our collisional results from \citet{dur2004} were obtained by tracking the results of impacts between two asteroids using the 3-dimensional SPH code SPH3D \citep{ba1994}. Computational details of these simulations can be found in \citet{dur2004,dur2007}. Once the ejecta flow field from the impacts are established, the results were handed off to the $N$-body code {\tt pkdgrav} 
\citep[e.g.,][]{rich2000,lei2000,stad2001,lr2002}. {\tt pkdgrav} is a scalable, parallel tree code for modeling the gravitational interactions between the resulting fragments. It has the ability to detect and treat low-speed collisions between particles, and allows rubble pile accumulations to form among ejected fragments. 

The target asteroids from \citet{dur2004} were 100-km-diameter undamaged basalt spheres. The spherical basalt projectiles had diameters of 10 to $46$~km: 10, 14, 18, 25, 34, and $46$~km, impact speeds from 2.5 to $7$ km~s$^{-1}$, and impact angles range from $15^\circ$ to $75^\circ$ (i.e., nearly head-on to very oblique). Details of the simulation outcomes are presented in Table~1 of \citet{dur2004}. We note that Erigone is a carbonaceous chondrite asteroid, and therefore may not be a good match to the target properties of basalt.  On the other hand, numerical impact experiments show that carbonaceous chondrite-like bodies, with lower densities but higher porosities than ordinary chondrite-like bodies, require similar collisional energies to produce catastrophic disruption events \citep[e.g.,][]{jut2015}. Given that our work is mainly an attempt to glean insight into the size of the Erigone parent body, we will stick to using the \citet{dur2007} results for now. Future work after Lucy's encounter with DJ can revisit this issue when more is known about the physical properties of DJ.

There are two observed SFDs for the Erigone family shown in the lower left panel of Fig.~\ref{fig2}. One uses WISE diameters (blue line), while the other converts the absolute magnitude of Erigone family members into diameters using the family's median albedo.  We tested both using the fitting procedure discussed in \citet{dur2007}. Our primary diagnostic values are the diameter ratio of the largest and the second largest remnants and the shape of the continuum SFD for bodies smaller than the second largest remnant. The model SFDs were scaled to match these values as best as possible, with the scaling factor telling us how much larger or smaller than parent body was likely to be from the starting size of $100$~km.  

Curiously, we were unable to find any satisfying fits to the WISE diameter SFD using the \citet{dur2007} model (Fig.~\ref{fig7}). We suspect this is because the largest two bodies have irregular shapes, making their WISE diameters less representative of their true diameters than one would expect. It could also be that the physical properties of basalt are not a good proxy for how carbonaceous chondrite-like material behaves in a disruptive collision. More work is needed to better understand this issue.

For the other Erigone family SFD based on the absolute magnitude of the family members and the family's median albedo, we found two reasonable fits to the \citet{dur2004} results.  They are defined as "Basalt\_5\_45\_1.8" and "Basalt\_6\_30\_2.2". Both are shown in Fig.~\ref{fig7}, and can also be found in Fig.~3 of \citet{dur2007}. The suffixes in the runs correspond to impact velocity, impact angle, and the logarithm of the target-to-projectile mass ratio \citep[see][for details]{dur2007}. Thus, for example, model Basalt\_5\_45\_1.8 involved a 100-km-diameter solid basalt target, impacted at $5$ km~s$^{-1}$ at a $45^\circ$ impact angle by a 25-km-diameter projectile.  

Overall, the fit between the two models is reasonable. We find that the red curve (Basalt\_5\_45\_1.8) provides a marginally better fit than the blue curve (Basalt\_6\_30\_2.2) because the red curve stays near or above the observed SFD. With that said, some of this is in the eye of the beholder; neither the red nor the blue curve fit the gold curve near $D = 15$~km. This mismatch could stem from interlopers, collisional evolution in the family over its lifetime, a poor estimate of the diameters of these bodies, the use of basalt rather than carbonaceous chondrite-like material, and so on. A closer inspection at the large members of the Erigone family does indicate that the reported error on their albedos is compatible with that of Erigone. We note, however, that these objects could be elongated, in which case the reported diameter would be biased toward larger size depending on observational geometry. The reader should also keep in mind that the SPH runs have a resolution limit, so all of the mass has to go somewhere, usually into the smallest particles. This behavior explains why the SPH SFDs become steep for $D < 2$~km.

Both model fits suggest that the Erigone parent body size was close to $80$~km in diameter. Calculating the mass remaining in the model fragment SFDs, and comparing that to the model parent body sizes, we predict that the Erigone family was produced by an impact modestly larger than a barely catastrophic disruption, where catastrophic means that $> 50$\% of the mass was ejected away at escape velocity. Both simulations show that roughly 60\% of the mass was put into family fragments. This would rule out the hypothesis that the Erigone family was created in a large cratering event.  

These values should be considered preliminary estimates for the properties of the Erigone family forming event. Given the current inability of the \citet{dur2007} model to match the SFD from WISE observations, additional work will be needed to see if this is a byproduct of an inadequate model or an interpretation issue affecting the observational data.   
\smallskip

\noindent{\it Collisional age of the Erigone family. }Immediately after the Erigone family was created, the family members began to be hit by asteroids from the background main belt population, which is nearly 3 orders of magnitude larger than the Erigone family itself \citep[i.e., one can compare the Erigone family SFD in Fig.~\ref{fig2} to the main belt SFDs shown in Fig.~1 of][]{bot2020}. Over time, these impacts will break down the Erigone family SFD, such that it will take on the same shape as the main belt SFD 
\citep[e.g.,][]{bot2005bel,bot2007,betal2015,broz2024a,bro2024}. Ideally, the shape change can be used like a clock to determine the age of the Erigone family, provided it is old enough that its observed SFD has had sufficient time to be noticeably affected by collisional evolution. The diagnostic constraint would be find a portion of the family SFD with a shape that is congruent with that of the main belt SFD.    

We decided to investigate this issue using the Collisional and Dynamical Depletion Evolution Model (CoDDEM) described in \citet{bot2005bel,bot2005a,bot2005b}. This code was used to track the collisional evolution of the main belt SFD over its history. Within this 1-D code, \citet{bot2005a,bot2005b} simulated how impacts changed the number of objects in a set of diameter bins (i.e., lower limit $D$, upper limit $D + dD$) between $0.0001 < D < 1000$~km, with logarithmic intervals set to $d\log D= 0.1$. Starting assumptions and computational details are provided in those papers. 

The most recent CoDDEM formulation of the main belt SFD can be found in \citet{bot2020}. They used it to model crater SFDs on spacecraft-observed asteroids like Ceres, Vesta, Lutetia, Mathilde, Ida, Gaspra, and Eros, all which are $D > 10$~km. Their crater SFDs have wavy shapes and spanned sizes between 0.1 and 100 km. Assuming that the crater to projectile ratio is $\approx 10$, as argued in \citet{bot2020}, these sizes correspond to asteroid diameters between 0.01 and 10 km (see below for more details about crater scaling laws). Accordingly, these data make it possible to constrain the shape of the main belt SFD well below the observational limit. Given that their preferred model main belt SFD provided an excellent match to these crater SFDs, one can make a case that the CoDDEM model and results from \citet{bot2020} can be used to simulate the collisional evolution of the Erigone family to reasonable approximation. This will be accomplished by (i) inserting an estimate of the initial Erigone family SFD into CoDDEM, (ii) tracking what happens to it over different family ages ($T$), and (iii) comparing the results to the observed Erigone family SFD.     

Our choice for the observed Erigone family SFD will come from Fig.~\ref{fig2}.  We will use the red curve, namely the SFD constructed using the absolute magnitudes and the median albedo of Erigone family members. The shape of the initial family SFD was determined through the following process. First, we tested how the observed SFD experienced collisional evolution over a range of $T$  values between $100-500$~Myr. We found that most collisions mainly affected the number of $D < 8$~km asteroids in the family SFD by a factor $f$, but only modestly affected the overall shape of the SFD. This meant that we could multiply the $D < 8$~km objects by $f$ and more or less reproduce the observed Erigone family SFD in time $T$. 

Second, we extrapolated the observed cumulative Erigone family SFD between $2 < D < 4$~km to smaller fragment sizes with $D < 2$~km. The cumulative power law slope found for this extrapolation was $q = -2.76$. For this preliminary exercise, we avoid using the hydrocode models from the last section, mainly because resolution issues mean the smaller sizes in the fragment SFD have more mass than they should in reality.  We should also state as a caveat that we do not know whether our assumption of a power law SFD for $D < 4$~km is valid, only that we do not have a better approximation.

\begin{figure}[t!]
 \begin{center}
 \includegraphics[width=0.49\textwidth]{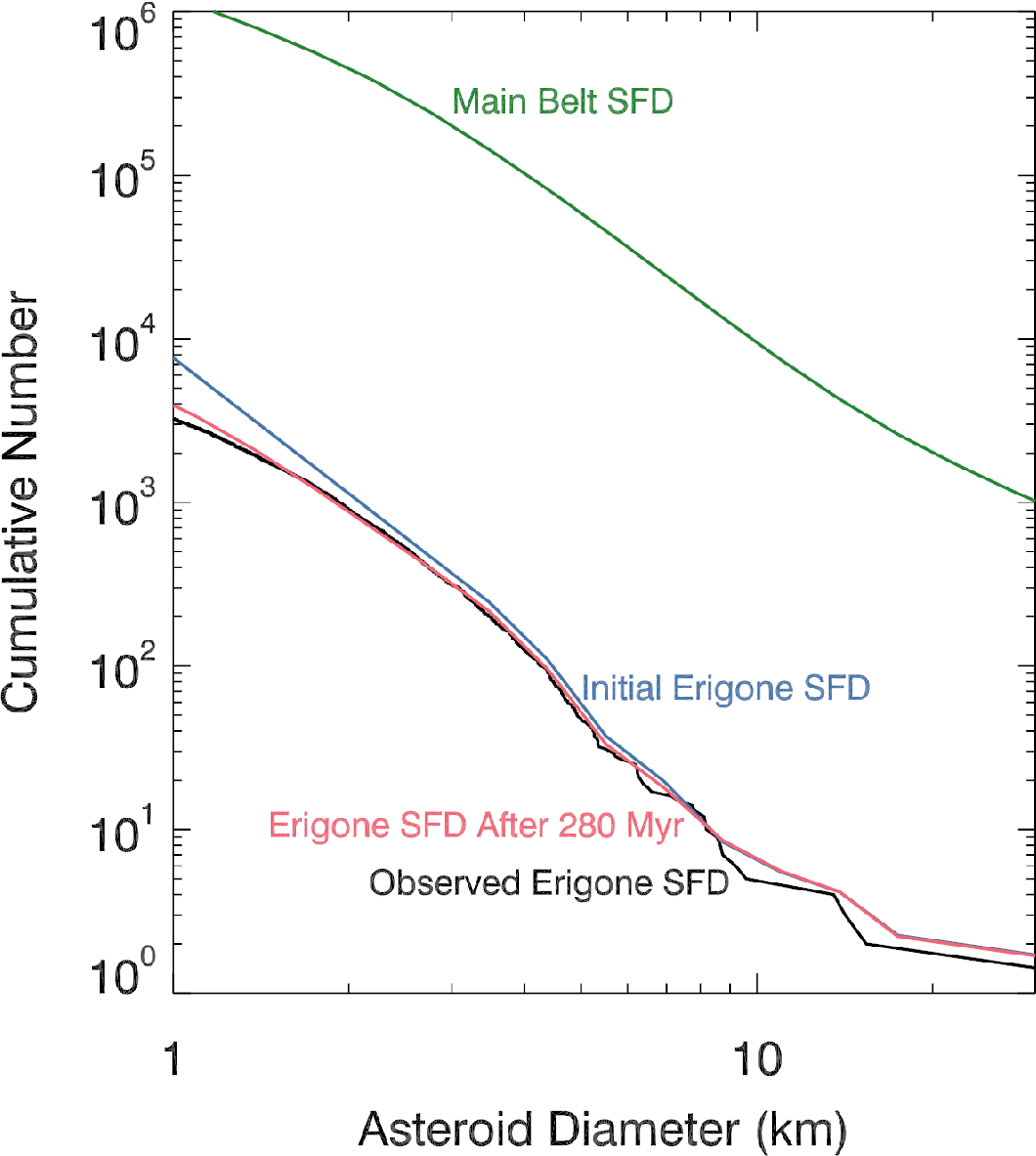} 
 \end{center} 
 \caption{Collisional evolution of the Erigone family SFD. The green curve represents the main belt SFD as formulated by \citet{bot2020}. The blue curve is our estimate of the initial Erigone family SFD. The black curve is the observed Erigone family SFD determined using absolute magnitudes and the family's median albedo (red curve in Fig.~\ref{fig2}).  The red curve shows the Erigone family SFD after $280$~Myr of collisional evolution.}
 \label{fig8}
\end{figure}

Finally, we used both components to test various $T$ values, with the goal to reproduce the observed SFD for $D < 2$~km. After some trial and error, we found the results that are shown in Fig.~\ref{fig8}.  The green curve is the cumulative main belt SFD from \citet{bot2020} (their main belt formulation~6), the black curve is the observed cumulative SFD, the blue curve is our estimate of the initial Erigone family SFD with $f = 1.2$, and the red curve is what the model SFD looks like after $280$~Myr.  Overall, the shape of the observed SFD is reproduced. There is a small mismatch near $D \approx 1$~km, but we suspect this is from observational incompleteness in the observed Erigone family. 

The reader should consider these results to be preliminary and approximate. We believe they should be revisited after Lucy's encounter with DJ, when we will know more about the physical properties of a real Erigone family member. With that said, these results help substantiate the dynamical ages suggested earlier in the text, namely that the Erigone asteroid family is $T < 300$~Myr old. Finally, we note that a typical 4~km main belt asteroid has a collisional lifetime exceeding 1~Gyr (Fig.~1 in \cite{mar06}), thus it likely DJ has retained the overall bulk properties since its formation.

\subsection{Does DJ belong to the Erigone family?} \label{djeri}
At this time, DJ's membership in the Erigone family is mainly based on circumstantial evidence. Given the available information, we believe that the pros outweigh the cons. We list some of our arguments below.

First, the number of interlopers in the Erigone family has been found to be low, especially for $a_{\rm P}>a_{\rm c}$ away from the $z_2$ secular resonance. DJ's $a_{\rm P}\simeq 2.384$~au meets this condition,
and avoids interaction with any meaningfully strong mean-motion resonance. 

As for DJ's orbital location, we note its high $e_{\rm P} \simeq 0.214$ value, which helps to explain its short orbital Lyapunov timescale of $\simeq 12$~ky. Tracking its orbits forward for $10$~Myr, we find DJ's
perihelion distance may reach $\simeq 1.727$~au. While this is not close enough to have a close encounter with Mars, it can approach this planet within several tens of its Hill radius. This leads to weak chaotic behavior in its orbit, a phenomenon termed ``stable chaos'' by \citet{mn1992}. 

This result prompted us to test what would happen to DJ over longer timescales. Here we used {\tt swift}, a well tested numerical package dedicated to integration of $(N+M)$-body problem (the Sun and $N-1$ massive planets plus $M$ massless particles), to propagate DJ's nominal orbit and 50 close clones forward in time for 1 Gyr. We included all eight planets and used a short timestep of $3$~days. We output our results every $5$~ky. We found that none of our DJ test asteroids was eliminated over the $1$~Gyr timespan. The instability timescale in DJ's orbital zone is therefore comfortably longer than the estimated age of the Erigone family. 

We also found that the proper value of the semimajor axis $a_{\rm P}$ of the clones experienced a random walk. In $200$~Myr,  their values were distributed in the interval $(-1,+2)\times 10^{-3}$~au about the initial value. This range is about an order of magnitude smaller than the expected semimajor axis changes caused by the Yarkovsky effect (see Fig.~\ref{fig6} for a rough estimate of the Yarkovsky shift of DJ over the best-fit Erigone family age $155$~Myr). It is therefore unlikely that large asteroids from the Erigone zone where DJ is located have leaked into to the terrestrial planet region. 

Note that the situation may be different for Erigone members having similar $e_{\rm P}$ values as DJ but $a_{\rm P}\leq 2.35$~au (see top and left panel on Fig.~\ref{fig1}). Here the bodies may be assisted by some weak mean motion resonances such as (J9,-S6,-2) at $a_{\rm P}\simeq 2.35$~au or J10/3 at $a_{\rm P}\simeq 2.33$~au.
 
Second, the prograde rotation and small obliquity value for DJ appear consistent with expectations based on our Yarkovsky/YORP evolution model (see also the Appendix~\ref{apper4} for context). DJ's $a_{\rm P}\simeq 2.384$~au represents the
transverse velocity difference of nearly $60$ m~s$^{-1}$ with respect the family center $a_{\rm c}$.
Our typical Yarkovsky/YORP evolution solutions presented in Sec.~\ref{family_now} resulted in $v_5\simeq
20$ m~s$^{-1}$. This means it is unlikely that DJ was initially ejected to its current orbital location. Rather, at least half of its semimajor axis distance from the family center was acquired by past Yarkovsky evolution. For that assertion to be true, DJ's past drift rate in $a_{\rm P}$ must have been positive, which requires prograde rotation. The independent determination of its rotation pole by recent telescopic observations (Mottola, 2024, pers. comm.) is therefore supportive to this model. DJ's slow rotation rate implies only a small lag between heating its surface by sunlight and thermal re-emission. As a consequence, the Yarkovsky effect is not optimum for its size. Indeed, DJ's location in the $(a_{\rm P},H)$ plane (bottom and left panel of Fig.~\ref{fig1}) stays away from the ``wavefront'' at the $C_\star$ isoline. Given our best guess parameters for DJ, we estimate a semi-major axis drift rate to be $da/dt \simeq 8.4\times 10^{-5}$ au~Myr$^{-1}$. Thus, the time scale to drift from the center of the family at $a_{\rm c}$ is about $172$~Myr (see Fig.~\ref{fig6} for more details).

Third, we consider DJ's slow rotation rate. The question is whether DJ's current rotation period of $P\simeq 252$~hr is compatible with an initial value of $P_0\leq 24$~hr followed by YORP evolution over a time interval constrained by the Erigone family age. We do not have the YORP effect measured for this asteroid, but we take the value detected for (1620) Geographos as a plausible template. Like DJ, Geographos is a very elongated object with a similar obliquity value and YORP rotation rate acceleration $d\omega/dt = (1.14\pm 0.03)\times 10^{-8}$ rad~d$^{-2}$
\citep[e.g.,][]{dur2022}. 

Considering that the strength of YORP scales with size $D$, semimajor axis $a$ and bulk density $\rho$, $d\omega/dt \propto \rho^{-1} (Da)^{-2}$, we obtain an estimate $d\omega/dt \simeq 2\times 10^{-9}$ rad~d$^{-2}$ for DJ (for simplicity, we take $10^{-9}$ rad~d$^{-2}$). Note that Geographos' rotation rate is accelerating by YORP, while here we assume the opposite evolution took place for DJ. Assuming that YORP has been decelerating DJ's rotation at the constant rate estimated above, and denoting $\kappa= 2\pi/P/(d\omega/dt)\simeq 1.64$~Myr, we estimate the required time $T$ to reach the current rotation period $P$ from its initial value $P_0$ as $T\simeq \kappa\, (P/P_0 -1)$. Taking $P_0\simeq 6$~hr, we get $T\simeq 70$~Myr. This value is shorter than the estimated age of the Erigone 
family, making it plausible that DJ formed from the Erigone family-forming event.  

Similarly, the YORP torque is able to change DJ's obliquity from a generic prograde initial state to near zero value on a timescale of approximately equal to the above-estimated $T$ for rotation-rate evolution.%
\footnote{Equations (3) and (5) in \citet{cv2004}, together with a charateristic YORP torque $T_\epsilon/C\propto -\sin\epsilon$ for prograde obliquities $\epsilon$, provide an approximate solution $\tan[\epsilon(t)/2]=\tan(\epsilon_0/2)\,(1-t/T)^\alpha$ with the initial value $\epsilon_0$, the timescale $T=\omega_0/(d\omega/dt)$ as in the main text, and the power exponent $\alpha=T\, (d\epsilon/dt)_0$ ($\omega_0$ is the initial rotation rate and $(d\epsilon/dt)_0$ the maximum obliquity rate for an asteroid rotation at frequency $\omega_0$). From data in Sec.~4 of \citet{cv2004} we estimate $(d\epsilon/dt)_0\simeq 1^\circ$~Myr$^{-1}$ and thus $\alpha\simeq 1.2$.}

An alternative possibility to the evolution of the rotation state by the YORP torques described above is that of a minimum amount of evolution. In that case, the current rotation would reflect the initial state at the Erigone family formation. Consider, for instance, that the DJ rotation is in fact in a tumbling state that would not be that surprising given its slow rotation \cite[see, e.g., Fig.~8 in][which indicates DJ is in the midst of parameter space populated by detected tumblers]{pra2014}. The uncertainty of the corresponding damping timescale is dominated by the unknown internal dissipation rate parameters, namely a product of the elasticity modulus $\mu$ and quality factor $Q$. But even in an optimistic situation $\mu Q\simeq 10^9$~Pa, DJ's tumbling requires more than hundred Myr timescale to become damped \citep[see Sec.~5.1 of][]{pra2014}. Note also that the tumbling rotation state of DJ would not conflict with its semimajor axis secular drift by the Yarkovsky effect. This is because the Yarkovsky effect has been both predicted and detected for a number of tumbling near-Earth asteroids, including (99942) Apophis or (4179) Toutatis \citep[e.g.,][]{vetal2015,far2024}.

\begin{figure}[t!]
 \begin{center}
 \includegraphics[width=0.49\textwidth]{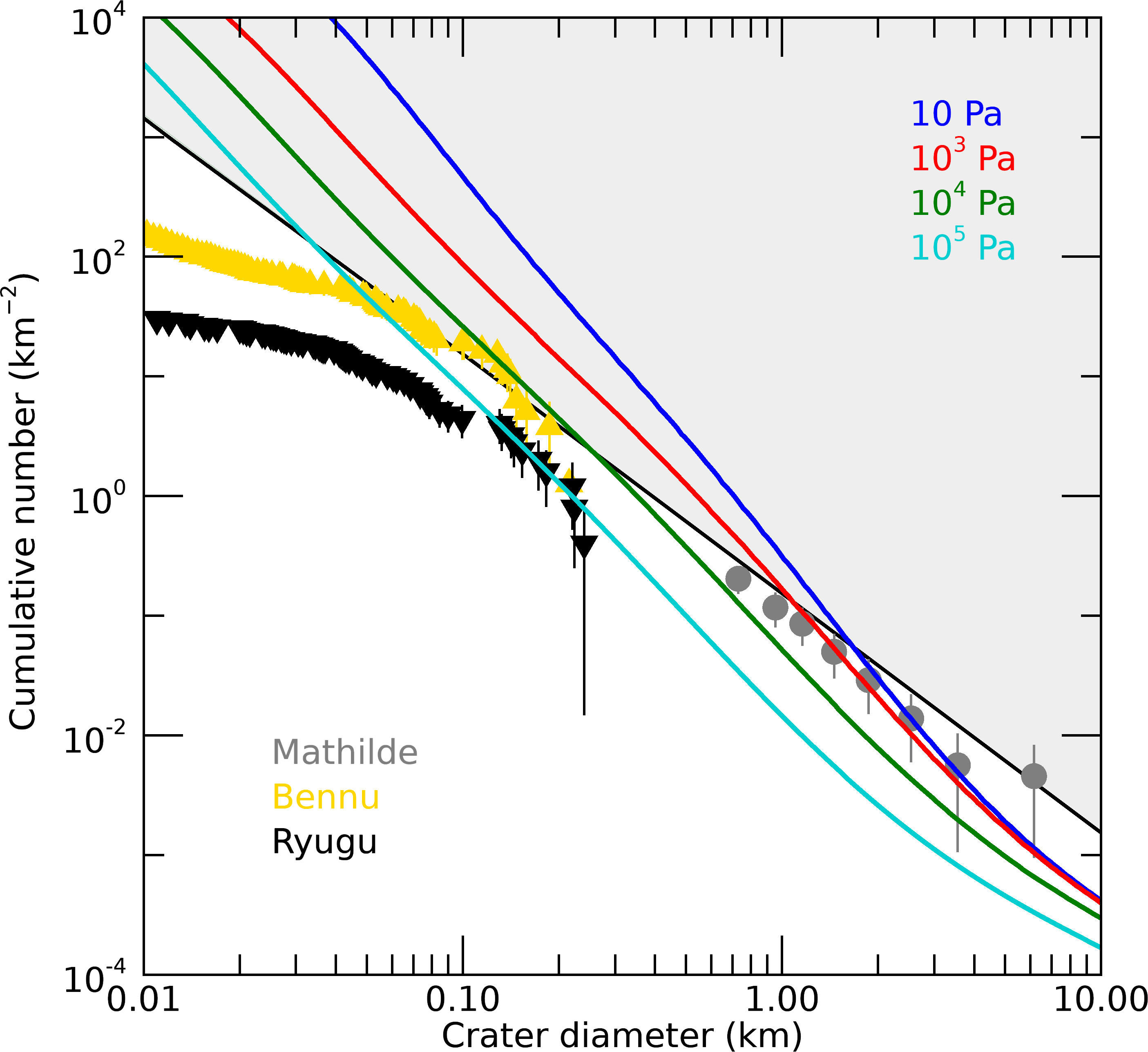} 
 \end{center} 
 \caption{Solid curves indicate our model crater SFDs for various target strength values $Y$ (as shown by the labels). The black line corresponds to the empirical crater surface saturation, as 10\% of the geometric saturation \citep{mar2015}. Observed crater SFDs for Bennu, Ryugu and Mathilde are given for sake of comparison \citep[yellow, green and gray, respectively; data from][]{bot2020,chap1999}.}
 \label{fig2s}
\end{figure}

\section{Implications for Lucy mission flyby} \label{lucy}
Here we explore the collisional history of DJ and make predictions on its surface crater SFD. These studies will serve as a reference and will support further analyses of the Lucy mission data.

For craters on DJ, we used the Pi-group scaling law \citep[e.g.,][]{hh2007a,hh2007b} that provides
the transient crater diameter as a function of impact conditions and material properties.  We 
further assume the final crater to be $\sim 30$\% larger than the transient crater
\citep[e.g.,][]{mar2015,mar2023}.

We calculated median values of the intrinsic collision probability $P_i$ and impact velocity $V_i$
using a sample of main belt asteroids larger than $50$~km \citep{fd1992} and obtained $P_i = 3.94\times 10^{-18}$ km$^{-2}$~yr$^{-1}$ and $V_i = 5.14$ km~s$^{-1}$. Figure~\ref{fig2s} shows
the computed crater cumulative SFD for DJ assuming a $150$~Myr surface age corresponding to the
best-fit family age. We implemented a cohesive soils cratering scaling law with various values of
target strength ($Y$), namely $Y = 10$, $10^3$, $10^4$, and $10^5$~Pa.

We stress that the strength of the target material is not known, and our assumptions provide a
reasonable range of properties. For instance, \citet{per22} found that the Bennu
surface strength is less than $2$~Pa, based on the ejecta pattern of a $70$ m diameter crater. On
the other hand, \citet{bal2020} concluded that meter-sized boulders on Bennu have a strength
of $0.5-1.7$~MPa. These strength values, however, are inferred from meter-scale properties
and their applicability to larger craters on DJ is not clear.

Our calculations show an interesting result, namely the crater SFD in the size range observable
by Lucy ($>100$~m in diameter) is sensitive to the material strength with shallower slopes for
increased strength $Y$ in the range $10-10^5$~Pa. When compared to a simple model for crater
surface saturation \citep[e.g.,][]{mar2015}, we find that DJ craters smaller than $1-2$~km could
be saturated even for the assumed young surface age for terrain strength $Y < 10^4$~Pa. For higher
$Y$ values, the computed crater SFD drops below saturation. Figure~\ref{fig2s} also shows the crater SFDs of 
Bennu, Ryugu and Mathilde (all C-types) for comparison. Note that Bennu ($\simeq 0.49$~km size) and
Ryugu ($\sim 0.90$~km size) are significantly smaller than DJ, while Mathilde ($\sim 53$~km size)
is significantly larger than DJ. Therefore, DJ fills a gap of asteroid size as concerns the craters
previously explored and measured by spacecraft.

\begin{figure}[t!]
 \begin{center}
 \includegraphics[width=0.49\textwidth]{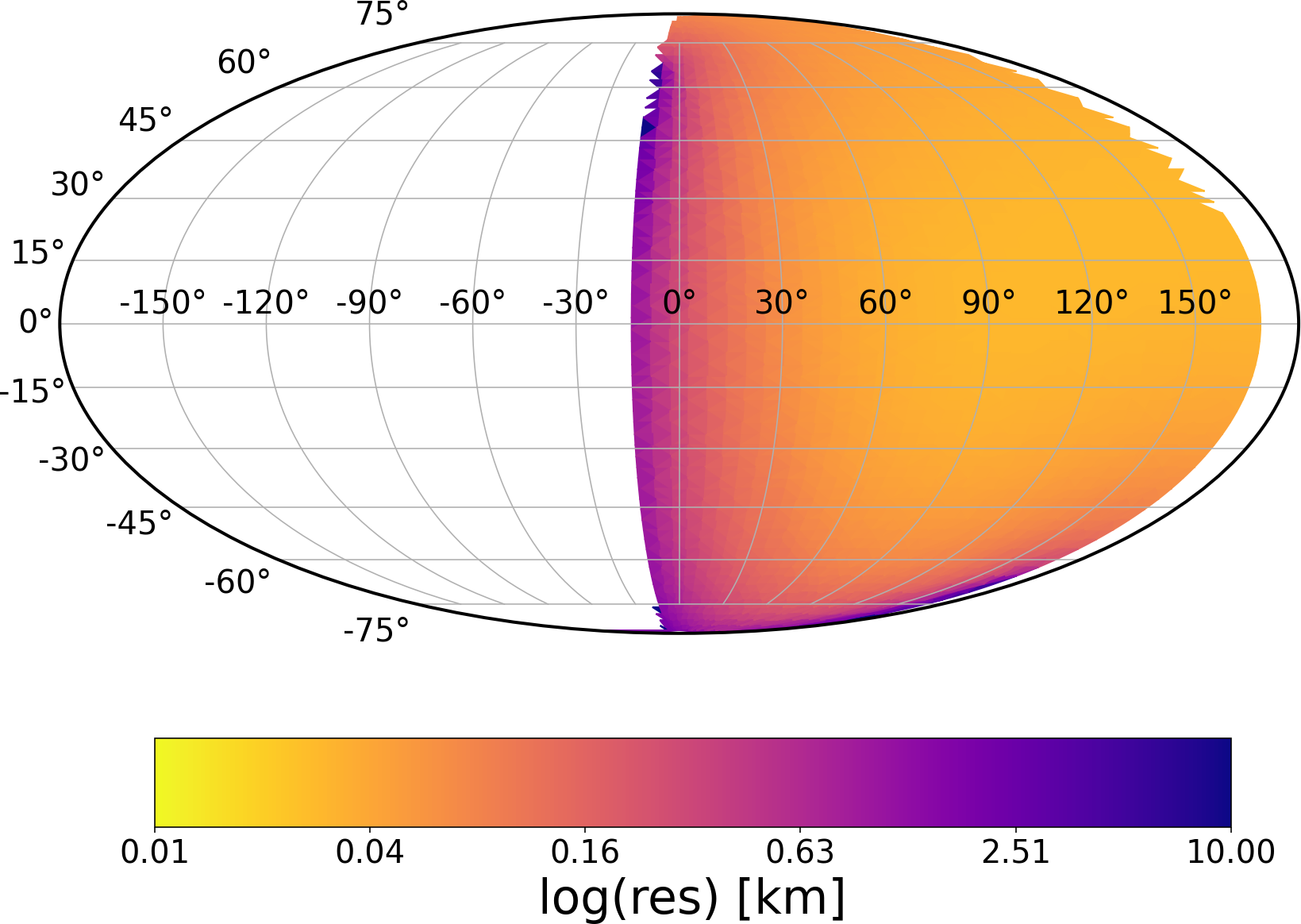} 
 \end{center} 
 \caption{Expected DJ surface coverage from Lucy flyby. The best resolution is about $10$ meters per
  pixel \citep[the term res indicates a resolution element, and corresponds to $3$ L'LORRI
  pixels]{wea2023}. Craters as small as $100$~m in diameter could be resolved on about one third
  of the imaged surface. However, we stress that this image assumes a spherical target, which may not
  be realistic.}
 \label{fig3s}
\end{figure}

The Lucy mission will flyby DJ and observe its surface with the high resolution L'LORRI imager
\citep{wea2023}. These observations are, however, limited by the fact that all observations will
terminate $\sim 40$ seconds before the closest approach distance due to solar elongation avoidance
constraints. As a result only 50\% of DJ’s surface will be imaged. The anticipated coverage and
spatial resolution is shown in Fig.~\ref{fig3s}. It is expected that Lucy will provide data in
a crater size range between $100$~m and a few km (depending on DJ actual size), which is a size
range not covered by Bennu/Ryugu and Mathilde.

\section{Conclusions}
In this paper, we presented our current understanding of the formation and evolution of the Erigone asteroid family and its member DJ, a flyby target of the NASA Lucy mission (slated for April 20, 2025). Due to a richness of available data regarding Erigone's family members, Erigone is an exemplary family for detailed studies about collisional formation and evolution. We show that the Erigone family is young ($<$250 Myr) and its fine orbital structure allows us to quantitatively constrain the physical parameters of DJ (e.g., density, thermal inertia).

The intention of this paper is to push the predicting potential of our dynamical and collisional models to the best of our capabilities, and then use Lucy flyby data to test our predictions. Given DJ inferred shape, spin state, composition, and collisional evolution, the DJ flyby offers the unique opportunity to test a wide range of model predictions.

In addition, the Erigone family is close to a cluster of primitive families in the inner main belt, from where Bennu and Ryugu originated. This offers a unique opportunity to study at close range a larger sibling in the native main belt environment before being pushed to near-Earth space or being shattered to smaller fragments by a catastrophic collision.

Finally, a note on the broader interest of the DJ flyby. The Lucy science team named asteroid DJ in honor of the discovered of the 3.2 million years old Lucy hominin (and namesake for the Lucy mission), paleoantropologist Donald Johanson.  As such, this is the first case in the history of space exploration in which a spacecraft visits an asteroid named after a contemporary human.

\acknowledgments
SM, WFB, HFL acknowledge support from the Lucy mission, financed through the NASA Discovery programme on contract no. NNM16AA08C. We thank M.~Bro{\v z} for providing us with a compilation of WISE albedo data. The work of DV and JD was partially supported by the Czech Science Foundation (grant 23-04946S).


\newcommand{\SortNoop}[1]{}

\appendix

\section{More details about the Erigone family}

\subsection{Determination of the formal family center and critical borderline in $C$ parameter}\label{apper1}
In order to make selection of the exterior interloper region in the $(a_{\rm P},H)$ plane objective
(Fig.~\ref{fig1}), we recall definition of the $C$-parameter using Eq.~(\ref{cline}). The sought
family limit is one of the $C$-isolines with a fine tuned value $C_\star$. The method of finding
$C_\star$, as well as the formal family center $a_{\rm c}$ was
originally proposed by \citet{wal2013}, and further developed by \citet{bol2017} and \citet{bol2018}.

Since most of the asteroid families --including Erigone-- do not exhibit huge asymmetries in distribution
of their members
in the $(a_{\rm P},H)$ plane (unless located close to major mean motion resonances), $a_{\rm c}$ is
typically located very near the largest asteroid in the family (in our case this would be at the
proper semimajor axis of (163)~Erigone). However, to cope with a slight degree of asymmetry, we
rather adjust $a_{\rm c}$ in order to match distribution of the bulk of smaller members in the family.
The method proceeds as follows. Assume $C_1$ and $C_2$ are two equal-sign values of the $C$-parameter
defined in (\ref{cline}) and $|C_2|>|C_1|$. Let then $N(C_1,C_2;a_{\rm c})$ denote number of asteroids
in between these two isolines. We use $N$ in two neighbor zones of width $\Delta C$ (in practice we
implemented $\Delta C=3\times 10^{-6}$~au) adjacent to a $C$-isoline to seek the largest drop
in the population (see the bottom and left panel of Fig.~\ref{fig1} to note the zones up to which
Erigone members piled up by the Yarkovsky drift). To determine the population contrast of the two
neighbor interval to $C$, we thus define
\begin{equation}
 r_+(C,a_{\rm c};\Delta C)=\frac{N(C-\Delta C,C;a_{\rm c})}{N(C,C+\Delta C;a_{\rm c})}\; , \label{rcrit1}
\end{equation}
for $C$ positive, and
\begin{equation}
 r_-(C,a_{\rm c};\Delta C)=\frac{N(C+\Delta C,C;a_{\rm c})}{N(C,C-\Delta C;a_{\rm c})}\; , \label{rcrit2}
\end{equation}
for $C$ negative. As we do not expect a significant asymmetry in Erigone members distribution in the
$(a_{\rm P},H)$ plane, we define the total contrast function
\begin{equation}
 r(C,a_{\rm c};\Delta C)=r_+(C,a_{\rm c};\Delta C) \,r_-(-C,a_{\rm c};\Delta C) \label{rcrit}
\end{equation}
for $C$ positive. We now seek optimum values of $a_{\rm c}$ and
$C_\star=C$ to maximize $r(C,a_{\rm c};\Delta C)$. Searching $a_{\rm c}\in (2.366,2.374)$~au and
$C\in (1.2,2.0)\times 10^{-5}$~au, we found the best solution $a_{\rm c}=2.3695$~au and $C_\star=1.7
\times 10^{-5}$~au. These are the parameters used in the bottom and left panel of Fig.~\ref{fig1}
(vertical dashed line and solid grey $C$-isolines).
\begin{deluxetable*}{rlcccccccr}[t] 
 \tablecaption{\label{known_models}
  Erione family members with known rotation models.}
 \tablehead{
  \colhead{number} & \colhead{designation} & $P$ & \colhead{$(\lambda,\beta)_1$} &
  \colhead{$(\lambda,\beta)_2$} & \colhead{$(\epsilon_1,\epsilon_2)$} & \colhead{$D$} & \colhead{$p_V$} & \colhead{Taxonomy} & \colhead{Refs} \\ [-3pt]
  \colhead{} & \colhead{} & (hr) & \colhead{(degrees)} & \colhead{(degrees)} & \colhead{(degrees)} & \colhead{(km)} & \colhead{} & \colhead{} & \colhead{} 
 }
\startdata
   163 & Erigone         & $\phantom{32}16.14$ & $(303,-67)$ &     $-$     & $(160,-  )$ & $69.4$ & $0.027$ & Ch,T & 3,4,5 \\
   933 & Susi            & $\phantom{321}4.62$ & $(299,-13)$ & $(122,-15)$ & $(105,103)$ & $24.7$ & $0.027$ & T  & 3,4  \\
  1448 & Lindbladia      & $\phantom{32}10.97$ & $( 83, 49)$ & $(273, 38)$ & $( 45,48 )$ & $22.7$ & $0.025$ & $-$ & 3 \\
  2776 & Baikal          & $3252.41$           & $( 90,-39)$ & $(270,-27)$ & $(124,122)$ & $18.3$ & $0.033$ & Cgh & 3,5 \\
  5026 & Martes          & $\phantom{321}4.42$ & $(199, 51)$ & $( 14, 60)$ & $( 35,34 )$ & $ 9.0$ & $0.066$ & Ch & 1,2,3 \\
  5506 & Artiglio        & $\phantom{321}9.41$ & $( 50, 78)$ & $(245, 69)$ & $( 15,18 )$ & $13.1$ & $0.039$ & X  & 3,5 \\
  8612 & Burov           & $\phantom{321}6.14$ & $(225,-20)$ & $( 49,-10)$ & $(106,104)$ & $ 5.2$ & $0.079$ & $-$ & 5 \\ 
  9566 & Rykhlova        & $\phantom{321}8.57$ & $(151,-72)$ & $(328,-73)$ & $(159,167)$ & $ 9.3$ & $0.065$ & Ch & 3,4 \\
 10527 & 1990 UN1        & $\phantom{321}5.78$ & $( 51,-73)$ & $(240,-60)$ & $(158,155)$ & $ 8.1$ & $0.050$ & $-$ & 3 \\ 
 14355 & 1987 SL5        & $\phantom{321}6.63$ & $(260, 62)$ & $( 94, 58)$ & $( 31,29 )$ & $ 6.9$ & $0.057$ & $-$ & 3 \\ 
 15758 & 1992 FT1        & $\phantom{321}7.55$ & $( 77, 45)$ & $(259, 33)$ & $( 50,52 )$ & $ 9.4$ & $0.038$ & $-$ & 3 \\ 
 18595 & 1998 BR1        & $\phantom{321}6.02$ & $(158,-75)$ &     $-$     & $(170,-  )$ & $11.5$ & $0.053$ & $-$ & 3 \\ 
 18851 & Winmesser       & $\phantom{32}27.32$ & $(302, 24)$ & $(125, 19)$ & $( 68,69 )$ & $ 7.6$ & $0.058$ & $-$ & 3 \\ 
 24723 & 1991 TW8        & $\phantom{321}6.17$ & $( 32, 29)$ & $(220, 27)$ & $( 62,61 )$ & $ 4.2$ & $  -  $ & $-$ &  \\ 
 24837 & Msecke Zehrovice & $\phantom{3}155.76$ & $( 13,-66)$ & $(232,-72)$ & $(157,163)$ & $ 5.3$ & $  -  $ & $-$ &  \\ 
 30772 & 1986 RJ1        & $\phantom{321}8.06$ & $( 80, 55)$ & $(259, 41)$ & $( 40,44 )$ & $ 6.3$ & $0.092$ & $-$ & 3 \\
 41707 & 2000 UU55       & $\phantom{32}13.35$ & $(  6,-58)$ & $(185,-42)$ & $(142,138)$ & $ 5.7$ & $0.072$ & $-$ & 3 \\ 
 44942 & 1999 VM55       & $\phantom{321}4.56$ & $(107, 32)$ & $(284, 42)$ & $( 54,52 )$ & $ 5.0$ & $0.060$ & X  & 3,4 \\
 49859 & 1999 XB100      & $\phantom{32}12.58$ & $( 29,-50)$ & $(211,-57)$ & $(144,164)$ & $ 7.4$ & $0.055$ & Cgh & 3,4,5 \\
 55440 & 2001 TY85       & $\phantom{3}428.09$ & $( 53,-72)$ & $(234,-77)$ & $(165,164)$ & $ 4.9$ & $0.072$ & $-$ & 3 \\ 
 61815 & 2000 QZ189      & $\phantom{321}9.21$ & $(199,-21)$ & $( 20,-34)$ & $(117,118)$ & $ 5.2$ & $0.059$ & $-$ & 3 \\ 
 64771 & 2001 XM180      & $\phantom{32}22.25$ & $(179,-40)$ & $(345,-48)$ & $(134,135)$ & $ 6.5$ & $0.038$ & $-$ & 3 \\
 66325 & 1999 JF55       & $\phantom{32}13.91$ & $(136,-54)$ & $(292,-59)$ & $(148,146)$ & $ 5.6$ & $0.088$ & Xc & 3,4 \\
 91345 & 1999 JK36       & $\phantom{3}369.06$ & $(327, 60)$ & $(127, 81)$ & $( 25,14 )$ & $ 5.6$ & $0.055$ & $-$ & 3 \\
 98345 & 2000 SQ304      & $\phantom{321}8.02$ & $( 95,-65)$ & $(292,-70)$ & $(157,159)$ & $ 4.6$ & $0.063$ & Cg & 3,4 \\
153364 & 2001 QL         & $\phantom{321}6.80$ & $(196,-67)$ & $( 37,-63)$ & $(155,156)$ & $ 4.8$ & $0.059$ & $-$ & 3 \\ 
\enddata
\tablecomments{The third column provides the sidereal rotation period $P$, the fourth and fifth columns provide the ecliptic longitude $\lambda$ and latitude $\beta$ of the rotation pole (photometric data analysis of the main belt asteroids often results in $\sim 180^\circ$ degeneracy of $\lambda$, thence the two pole solutions in some cases), the sixth column provides obliquity $\epsilon$ of the pole solution(s), the seventh and eight columns give size $D$ and geometric albedo $p_V$ from WISE data (except in cases not observed by WISE for which the size is estimated from the absolute magnitude value and albedo ${\bar p}_V=0.057$), the ninth column gives the spectral class, and the tenth column provides principal references: 1 ... \citet{pra2019}, 2... \citet{pol2014}, 3 ... \citet{mas2011}, 4 ... \citet{mor2016}, 5 ... \citet{har2024}. The associated shape models, when available, and other information could be downloaded from the {\tt DAMIT} website \citep[see also][]{dur2010}.}
\end{deluxetable*}
\begin{figure}[t]
 \begin{center}
 \includegraphics[width=0.47\textwidth]{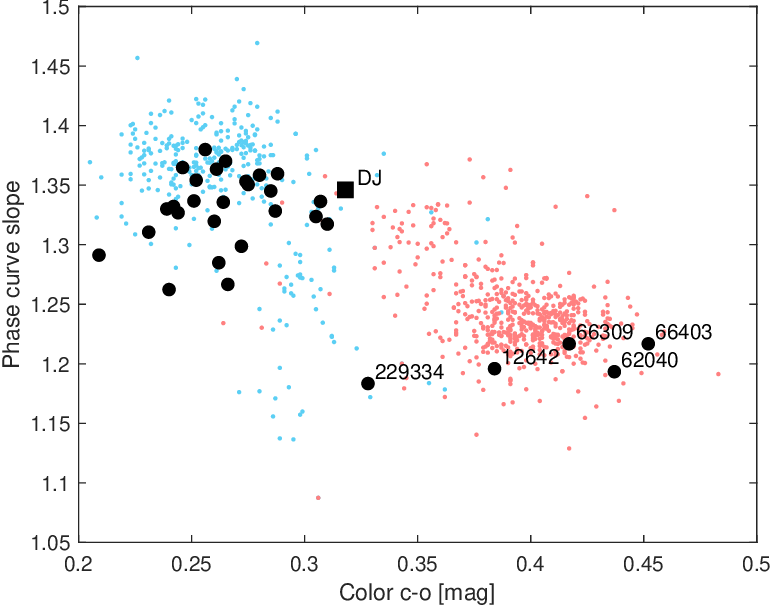} 
 \end{center} 
 \caption{Color index c-o defined as a difference between magnitudes in cyan and orange ATLAS filters (abscissa) and slope of the phase curve (ordinate) for asteroids which (i) are nominally members of the Erigone family, and (ii) for which the available photometric data allow us to determine their rotation state (i.e., sidereal rotation period and pole orientation). Distinct cluster of black points characterized by a small color index and large slope corresponds to Erigone family members with low albedo (these are listed in Table~\ref{known_models}). Importantly, DJ --shown in black square-- appears to belong to this group. Additionally, there are five asteroids, identified by their numbers, clearly separated from the main cluster by having larger color index and smaller slope values. These are likely interlopers with high albedo and affinity to S-taxonomic class. As a result, we disregard them in our further analysis. The underlying red and blue points are known asteroids of S and C taxonomic class projected onto these axes \citep[compare with Fig.~4 in][]{dur2020}.}
 \label{fig_color_slope}
\end{figure}
\begin{figure*}[t!]
 \begin{center}
 \includegraphics[width=0.95\textwidth]{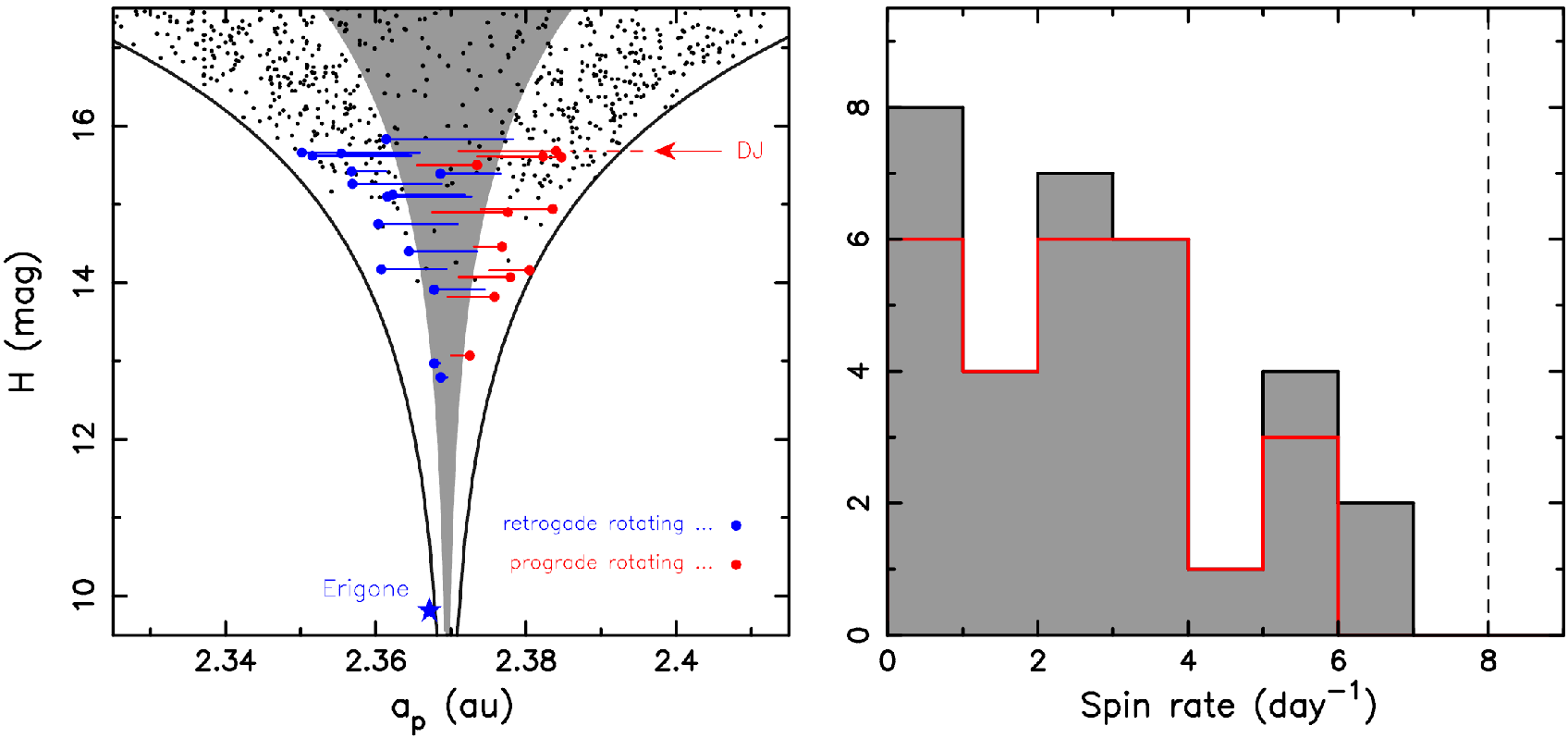} 
 \end{center} 
 \caption{Left panel: Segment of Erigone family projected onto the plane defined by the proper semimajor axis $a_{\rm P}$ (abscissa) and absolute magnitude $H$ (ordinate; $H< 17.5$~magnitude zone shown). The black symbols are Erigone members observed by WISE constrained by two conditions: (i) $p_V\leq 0.125$, and (ii) with $|C|\leq |C_\star|=1.7\times 10^{-5}$~au (grey lines). The light grey V-shaped zone is defined by $|C|\leq 5.2\times 10^{-6}$~au, and it corresponds to region in which the initial fragments from the family-forming event would land assuming isotropic velocity field and ejection velocity $v_5=24$ m~s$^{-1}$ for $D=5$~km asteroids \citep[see Eq.~(5) in][]{vok2006a}. The color-highlighted asteroids with known spin state, sidereal rotation period and pole orientations, consist of DJ and objects listed in Table~\ref{known_models}: the prograde- and retrograde-rotating cases are red and blue. The color intervals connect their current position to their past state $T=155$~Myr ago using a simplified steady migration by the Yarkovsky effect fixing the present-day rotation state, bulk density $\rho=1.5$ g~cm$^{-3}$ and surface thermal inertia $\Gamma=135$~SI (in the DJ case we assumed zero obliquity, but any value $\leq 20^\circ$ would lead to about the same conclusions). As expected, the prograde- and retrograde-rotating objects have $a_{\rm P}>a_{\rm c}$ and $a_{\rm P}<a_{\rm c}$, respectively, and their initial positions are in (or near) the initial-velocity dispersal zone. Right panel: Distribution of rotation rate values (abscissa; in rotations per day) for Erigone family members (except for (163) Erigone; grey histogram). The same for subsample of asteroids with resolved rotation pole orientation in Table~\ref{known_models} is higlighted in red. The distribution is not compatible with that of collisionally relaxed population, characterized by the Maxwellian function, rather it bears similarity to the YORP-relaxed sample \citep[see][for comparison]{pra2008}. The vertical dashed line is the approximate rotation fission limit for low-density and low material cohesion asteroids.}
 \label{fig6}
\end{figure*}

\subsection{Erigone population completeness at $H=17.5$ magnitude limit}\label{apper2}
We used observations by Catalina Sky Survey (CSS), Mt.~Lemmon survey telescope (IAU code G96),
in between January 2013 and July~2023. \citet{neomod2} performed a detailed analysis of
the detection efficiency of hundreds of thousands of fields-of-view taken by the telescope
during this periods of time, helping them to develop a new population model for the
near-Earth asteroid population. Here we use the same database of well-calibrated observations
with known sky-coverage to infer detection probability of orbits in the Erigone family zone.

First, we simply sifted all the CSS/G96 frames taken in the 2013/2023 decade and searched
for detections of the members in our nominal Erigone family (Sec.~\ref{family_now}). Out
of all 4925 members, CSS observed 4291 of them, missing just 634 asteroids. In the group of
unobserved members, only (20992)~Marypearse stands out with its $H=15.52$ magnitude. All
others, starting with (391495) 2007~OV10, have $H\geq 17.62$ and are thus small members
(some of which have been even discovered after July~2023). In fact only three of them have
the absolute magnitude $H<18$ (all of them having $a_{\rm P}> 2.38$~au).

Second, we created two samples of $15000$ synthetic orbits in the Erigone family by
considering their real orbits with semimajor axes in the range $(2.32,2.33)$~au (first
sample) and $(2.41,2.43)$~au. We assigned them their osculating orbital elements as
of MJD 60600.0 epoch, with the only exception of the mean anomaly that was randomized
in the $0^\circ$ and $360^\circ$ range. These orbits were propagated backward in time to
January~2013 with the goal to infer whether (i) they would fall in one of the CSS fields of
view, and (ii) if so, whether the telescope would detect them.  To that end, we use the publicly
available and well-tested {\tt objectsInField code} from the Asteroid Survey Simulator package
\citep{naidu2017}. We found that in the first sample all objects with $H\leq 17.3$ have
basically 100\% detection probability, and those with $17.5$ have 99.2\% detection
probability. Likewise, objects in the second sample have 100\% detection probability for
$H\leq 17.05$ magnitude, and even those with $H=17.5$ have 97.5\% detection probability.
Beyond magnitude $H=18$ the detection probability drops below 90\%, and there is approximately
$0.25$ magnitude shift in between the two samples to reach the same detection probability
(i.e., in the group with smaller semimajor axes objects with absolute magnitude larger by about
$0.25$ are detected with the same probability as those in the group with larger semimajor axes).
This bias is seen at the bottom and left panel of Fig.~\ref{fig1}.

From this exercise, taking into account only CSS observations, we infer near completeness of
the Erigone family at $H=17.5$ magnitude. This conclusion would be even strengthened, if
well characterized observations by other prolific surveys (sun as PanSTARRS) were available to
us. The result also agrees with that reached by an independent method in \citet{hen2020}.

\subsection{Erigone members with known rotation state}\label{apper4}
The Yarkovsky/YORP evolution model of the families, when applicable, implies a particular distribution of rotation state of their members. This is because Yarkovsky-driven migration to large or small values of the proper semimajor axis $a_{\rm P}$ requires prograde or retrograde rotation state. The analysis of sparse photometry, provided by powerful ground- and space-based surveys, recently allowed to significantly increase the sample of asteroids with resolved rotation state. Consequently, the Yarkovsky/YORP dynamical models for a number of asteroid families could have been eventually tested \citet[see][for several spectacular examples]{dh2023}.

While solutions for spin state of several Erigone members have been previously published, we decided to re-evaluate them with a critical eye on statistically borderline cases. To that goal, we collected sparse photometry from the main sky surveys \citep[for details, see][]{Han.ea:23} of Erigone members and used the lightcurve inversion method of \cite{Kaa.ea:01}. In some cases, we found previously published models not statistically robust enough, in several other cases we obtained new solutions. The final models are summarized in Table~\ref{known_models}. We note that the available photometric data allow to solve for rotation period and pole of five more nominal members in the Erigone family, namely (12642) Davidjansen, (62040) 2000~RA64, (66309) 1999~JX41, (66403) 1999~LM3, and (229334) 2005~QV4. However, analysis of their color indexes (see Fig.~\ref{fig_color_slope}) suggests they belong to high-albedo interloper class and, consequently, we discarded them from valid sample of Erigone asteroids with resolved spin state.%
\footnote{In the case of (12642) Davidjansen and (66309) 1999~JX41 this conclusion is also supported by broad band photometry data from SDSS, since their principal component value $a^\star > 0.1$ make them belong to the high-albedo S-type taxonomic group \citep[e.g.][]{par2008}.} As shown by \cite{dur2020}, the slope of the phase curve (defined as the ratio of the theoretical phase curve function at 10 and 20 degrees of the solar phase angle) and the color index c--o in ATLAS c and o filters are correlated with albedo -- low-albedo C-complex asteroids have higher slope and smaller difference between c and o filters, while high-albedo S-complex asteroids have smaller slope and larger difference between filters. Because all modeled Erigone family members have observations from the ATLAS telescopes in both filters\footnote{We used The ATLAS Solar System Catalog (SSCAT) Version 2 available at \url{https://astroportal.ifa.hawaii.edu/atlas/sscat/}.}, we can use this diagnosis tool for detecting high-albedo interlopers (Fig.~\ref{fig_color_slope}). 

Figure~\ref{fig6} (left panel) displays the results in the relevant projection to the plane of proper semimajor axis $a_{\rm P}$ and absolute magnitude $H$. As outlined above, we expect prograde rotating members (red symbols) located in the $a_{\rm P}>a_{\rm c}$ zone, while the retrograde rotating members (blue symbols) located in the $a_{\rm P}<a_{\rm c}$ zone. The largest members, or those with the rotation pole near enough to the ecliptic plane, may be located near the family center. In order to characterize what "the center" means for asteroids of a different magnitude value, we consider the formally best-fit Yarkovsky/YORP model from Sec.~\ref{family_now} that had $v_5=24$ m~s$^{-1}$ initial ejection velocity of $D=5$~km fragments (and $v_{\rm ej}\propto 1/D$ for other fragments). Their initial location is highlighted by the light-grey region near the family center (assuming the isotropic ejection field of the fragments). Finally, we consider the estimated value of the Yarkovsky drift $d a_{\rm P}/dt$ using the present-day obliquity and parameters of the best-fit model, in particular the $155$~Myr age of the Erigone family, and mapped the current to the initial asteroid location (see the color intervals). This procedure is obviously highly simplified, as no YORP modification of the rotation parameters are taken into account, but provides at least a zero order estimate of the asteroid evolution in these coordinates over the predicted family age. The right panel in Fig.~\ref{fig6} shows distribution of rotation rate values for Erigone fragments: the distribution for the most complete models, for which also the rotation pole has been determined (Table~\ref{known_models}) is highlighted in red, while the grey histogram is for all Erigone members \citep[period solutions for cases in which the rotation pole has not been determined were taken from the {\tt LCDB} database; see][]{war2009}.
In either case, the rotation rate distribution cannot be matched with a Maxwellian model for collisionally relaxed population. The surplus of slowly-rotating objects hints traces of a population evolved by the YORP effect \citep[see][for context]{pra2008}.

Overall the limited data provide an excellent justification of the model, since (i) the prograde- and retrograde-rotating members are all located in their respective zones, (ii) the estimated evolutionary tracks connect their position to the zone, where they should be located initially, and (iii) the rotation rate distribution reveals perturbation by the YORP effect. We expect this picture will be even strengthened when more solutions will be added in the future.

\subsection{Outline of the Yarkovsky/YORP chronology approach}\label{apper3}
The method of Yarkovsky/YORP chronology of the asteroid families has been developed in
\citet{vok2006a}, with a precursor version dwelling on the Yarkovsky component in \citet{nes2003}.
Since both \citet{vok2006a} and \citet{bot2015} provide a detailed description of the approach,
here we list the principal steps in brief (relegating the reader to those references for more 
information if needed). 

We assume the family has been represented using the differential distribution $dN(C)$ with the
$C$ parameter defined by Eq.~(\ref{cline}). In the case of the Erigone family, we determined the
family center $a_{\rm c}$ and maximum-contrast $C_\star$ parameter in the Appendix~\ref{apper1}.
In order to describe drop in the family population beyond $C_\star$, we extend to $dN(C)$ to
$C_{\rm fam}=1.95\times 10^{-5}$~au at which the family population is effectively nil. We use the
finite bin-size $\Delta C = 1.5\times 10^{-6}$~au to represent the distribution, and we use the
symmetric version $dN_{\rm sym}(C)$ defined in Eq.~(\ref{dnsym}). In order to minimize contamination by
interlopers and avoid problems with uncertain modeling of the YORP evolution over several of its
cycles (evolution from a generic initial conditions to the asymptotic state), we used 640 Erigone dark
members with $H\leq 17.5$ were used to construct these data.

The goal of the model is to match the observed-family $dN_{\rm sym}(C)$ distribution with prediction
$dM(C;{\bf p})$ by minimization of a target function (\ref{tg}). The minimization is achieved by
determining the optimum (best-fit) model parameters ${\bf p}$. We split ${\bf p}$ into two categories:
(i) the original set introduced by \citet{vok2006a}, and (ii) additional parameters adopted fixed
by \citet{vok2006a}, but given that their uncertainty also appreciably affects the results, we decided to
include them in the present work. The first set 
consists of (i) the family age $T$, (ii) the characteristic initial ejection velocity $v_5$ of $5$~km
size fragments (fragments of size $D$ are assumed to be ejected with velocity $v_5\,(5\,{\rm km}/D)$),
and (iii) the empirical YORP torques strength parameter $c_{\rm YORP}$, adjusting them against a simple
template taken from \citet{cv2004}. The second set includes (i) asteroids' bulk density $\rho$, (ii)
asteroids' surface thermal inertia $\Gamma$, and (iii) characteristic timescale $\tau_0$ to reset the
YORP strength for a kilometer size member of the family. Therefore ${\bf p}=({\bf p}_1;{\bf p}_2)=
(T,v_5,c_{\rm YORP};\rho,\Gamma,\tau_0)$. The meaning of the first five parameters is straightforward.
The last --$\tau_0$-- merits a brief explanation.

\citet{bot2015} developed what they called a ``variable (or stochastic) YORP'' approach which was found
to suit modeling the Eulalia family better than the traditional ``static YORP'' approach. In the latter
case the characteristic strength of the YORP torques is kept constant over asteroid's lifetime, while
in the former case, it is allowed to change on a characteristic timescale. This is because the magnitude
of the YORP torques was found very sensitive on the small-scale surface topography 
\citep[e.g.,][for a review]{vetal2015}. Therefore, formation of new craters and the related surface shaking,
which may cause landslides or boulder mobility, all may affect the YORP influence on small asteroids'
rotation-state evolution. \citet{bot2015} thus considered a characteristic YORP-change timescale 
$\tau_{\rm YORP}$ over which they changed the YORP-strength coefficient pre-computed by \citet{cv2004}.
Since the essence of the effect may have to do with small (sub-catastrophic) impacts, we additionally 
assume $\tau_{\rm YORP}=\tau_0\,\sqrt{D}$, with $\tau_0$ a free parameter and $D$ the size in kilometers
\citep[the power-exponent of the size dependence was motivated by analysis in][]{fvh1998}. In our production
simulations we consider $\tau_0$ in the $0.5$ to $100$~Myr range (note that $\tau_0\rightarrow \infty$ is the
formal limit to the static approach of YORP modeling). An important implication of the variable YORP is 
a separation of the evolution of the obliquity and rotation-rate timescales \citep[see][]{bot2015,s2015}:
while the obliquity evolves at the regular timescale of the static YORP variant, the rotation-rate evolution
is slowed down by effects of random walk. This helps the process of small asteroid piling up to the semimajor
axis extreme values in the family. 

The model itself operates in the $(a,H)$ plane in which it initially (at the family formation) creates a
synthetic population of 640 Erigone fragments centered about $a_{\rm c}$ and distributed in $a$. This model
population has the same sizes $D$ as the real Erigone members that contributed to construction of the 
$dN_{\rm sym}(C)$ distribution, each of them is ejected with velocity $v_5\,(5\,{\rm km}/D)$, and the initial 
velocity field is assumed isotropic. The fragments are also given initial rotation-state parameters, namely
(i) rotation period $P$ from a Maxwellian distribution peaked at $6$~hr, and (ii) obliquity $\gamma$
corresponding
to an isotropic distribution. Since the model fundamentally aims at describing the family dynamical evolution,
the time $t$ is let advance in steps $dt=0.1$~Myr. During this process, each fragments' $(a;P,\gamma)$ are
evolved: (i) semimajor axis $a$ by the Yarkovsky effect, and (ii) $(P,\gamma)$ by the YORP effect (details and
complexities of this modeling are described in \citet{vok2006a} and \citet{bot2015}). At regular timesteps
$dt'=2$~Myr, the synthetic family population is mapped to the $C$-space and the corresponding distribution
function $dM(C;{\bf p})$ determined and the target function (\ref{tg}) computed. 

\end{document}